**Evaluation of the performance of Euro-CORDEX RCMs for assessing hydrological climate change impacts in Great Britain: a comparison of different spatial resolutions and quantile mapping bias correction methods**

**Authors**

Ernesto Pastén-Zapata (1 and 3), Julie Jones (1), Helen Moggridge (1), Martin Widmann (2)

((1) Department of Geography, University of Sheffield, UK, (2) School of Geography, Earth and Environmental Sciences, University of Birmingham, UK, (3) Geological Survey of Denmark and Greenland (GEUS), Department of Hydrology, Denmark)

Keywords

RCM, Quantile mapping, CORDEX, climate change, hydrological modelling, HEC-HMS

**Abstract**

Regional Climate Models (RCMs) are an essential tool for analysing regional climate change impacts as they provide simulations with more small-scale details and expected smaller errors than global climate models. There has been much effort to increase the spatial resolution and simulation skill of RCMs, yet the extent to which this improves the projection of hydrological change is unclear. Here, we evaluate the skill of five reanalysis-driven Euro-CORDEX RCMs in simulating precipitation and temperature, and as drivers of a hydrological model to simulate river flow on four UK catchments covering different physical, climatic and hydrological characteristics. We use a comprehensive range of evaluation indices for aspects of the distribution such as means and extremes, as well as for the structure of time series. We test whether high-resolution RCMs provide added value, through analysis of two RCM resolutions, 0.44° (50 km) and 0.11° (12.5 km), which are also bias-corrected employing the parametric quantile-mapping (QM) method, using the normal distribution for temperature, and the Gamma (GQM) and Double Gamma (DGQM) distributions for precipitation. The performance of these is considered for a range of meteorological variables and for the skill in simulating hydrological impacts.

In a small catchment with complex topography, the 0.11° RCMs outperform their 0.44° version for precipitation and temperature, but when used in combination with the hydrological model, fail to capture the observed river flow distribution. In the other (larger) catchments, only one high-resolution RCM consistently outperforms its low-resolution version, implying that in general there is no added value from using the high-resolution RCMs in those catchments. GQM decreases most of the simulation biases, except for extreme precipitation and high flows, which are further decreased by DGQM. Bias correction does not improve the representation of daily temporal variability, but it does for monthly variability, in particular when applying DGQM, which reduces most of the simulation biases. Overall, an increase in RCM resolution does not imply a better simulation of hydrology and bias-correction represents an alternative to ease decision-making.

## 1. Introduction

Global General Circulation Models (GCMs) are the main tool for climate change projections. However, their spatial resolution is usually not higher than 100 km (Rummukainen, 2016), limiting their skill to simulate local climate. Regional Climate Models (RCMs) focus on specific subcontinental or subnational domains, incorporating regional features such as topography, coasts and islands more accurately. Consequently, RCMs improve the simulation of small-scale processes that affect precipitation, such as orographic forcing (Rummukainen et al., 2015; Di Luca et.al, 2015), and are expected to yield more accurate projections of climate change at finer spatial scales. RCMs have been used extensively to evaluate the impacts of climate change on hydrology, such as changes in mean river flow, floods or low flows (e.g. Kay et al., 2015; Kay and Jones, 2012; Mendoza et al., 2016; Teng et al., 2015; Prudhomme et al., 2013; Cloke et al., 2013).

The resolution of RCMs has increased over time with the availability of higher computer power. Currently, the spatial resolution of RCMs varies from 50 km to less than 5 km (Rummukainen, 2016; Rockel et al., 2015). Due to their increased representation of regional features and small-scale processes, RCMs generally simulate the current regional climate better than their driving data (Feser et al. 2011; Di Luca et al., 2015). Nevertheless, this might not be true in regions mainly influenced by large-scale climatic processes (Eden et al., 2014). Therefore, the added value of high-resolution RCMs depends on the analysed region, variable and context (Rummukainen, 2016).

An important driver for increasing RCM resolution is the need to improve the analysis of climate change impacts for decision-making (e.g. Macadam et al., 2016; Qian et al., 2015). For hydrology, the standard analysis of climate change impacts generally involves coupling uncorrected or bias-corrected GCM or RCM precipitation and temperature outputs with hydrological models to simulate river flow scenarios (e.g. Teutschbein and Seibert; 2012; Huang et al., 2014; Teng et al., 2015). In Great Britain, these studies focus on one (or more) of four main topics: 1) the contribution of the GCMs, RCMs, emission scenarios and bias-correction techniques to the uncertainty of the change projection (e.g. Prudhomme and Davies, 2009; Kay et al., 2009; Arnell, 2011; Christierson et al., 2012), 2) the impact of the bias correction techniques to the projections (e.g. Prudhomme et al., 2013; Cloke et al., 2013; Wetterhall et al., 2012; Kim et al., 2016), 3) projections of future floods (Cloke et al., 2013; Kay et al., 2015; Wetterhall et al., 2012; Kay and Jones, 2012), and, 4) projections of future low flows (Wilby and Harris, 2006; Arnell, 2011; Fowler and Kilsby, 2007).

Some studies have identified a consistent improvement in hydrological simulation skill with increasing RCM resolution for the annual mean river flow (Huang et al. 2014). For the simulation of river flow peaks as a response to extreme precipitation events, previous studies found no improvement when increasing the model resolution (Kay et al. 2015; Huang et al., 2014). Others studies found that the improvement depends on the catchment size and on the evaluation index (Dankers et al. 2007), whilst others found an improvement when simulating seasonal flow and hydrologic signatures aimed to represent diverse hydrologic processes (e.g. runoff ratio, center time of runoff) (Mendoza et al., 2016). However, these studies have only used one RCM to perform the comparison as, to date, there has been no systematic study using a large number of RCM simulations to test the effect of RCM resolution on hydrological simulation skill.

The first aim of this paper is to use the EURO-CORDEX simulations (Jacob et al. 2014) to robustly assess the added value of increasing RCM resolution on hydrological simulations. The Euro-CORDEX simulations at 0.11° (12.5 km) and 0.44° (50 km) have the same lateral boundaries and the parameterisations of each RCM are the same at both resolutions, thus making them ideal for such a comparison. This work builds on assessments of the 0.11° and 0.44° Euro-CORDEX RCMs at reproducing observed temperature and precipitation distributions, including extremes and dry/wet spell lengths. Results vary among the studies. Some found a higher accuracy for the 0.11 RCMs for Europe when evaluating the mean and extreme precipitation at a daily and sub-daily temporal resolution (Prein et al. 2015, Fantini et al 2016), whereas others did not find an improvement in accuracy when assessing the spatio-temporal patterns of the monthly and seasonal precipitation and temperature (Kotlarski et al. 2014). For the Alps Torma et al. (2015) found a higher skill for the 0.11 RCMs when simulating the spatial distribution of the mean, extreme and intensity of precipitation, while Casanueva et al. (2016) showed for the Alps and Spain that the best performance depends on the RCM, season and validation index when evaluating precipitation intensity, frequency, mean and extremes.

Biases in RCM simulations are due to parameterization of sub-grid processes, limited representation of local features, incorrect boundary conditions and differences between spatial resolutions of the simulations and observations (Ehret et al., 2012; Benestad, 2010). Therefore, RCMs require post-processing for many applications (Christensen et al., 2008). Statistical bias-correction techniques reduce biases in the mean, variance or the complete distribution of simulated climate variables (reviews in Maraun et al., 2010; Teutschbein and Seibert, 2012; Maraun and Widmann, 2018; Lafon et al., 2013). Quantile mapping (QM) is one of the standard techniques used (Piani et al., 2010; Teutschbein and Seibert, 2012; Maurer et al., 2014). Whilst effective, bias correction has important limitations that are further discussed in the conclusions.

To date, a detailed comparison of the simulation skill of bias-corrected high- and low-resolution model outputs for aspects that are important for hydrological studies (e.g. means, extremes, daily sequence) has not been undertaken. The second aim of this study addresses this research gap by conducting a detailed evaluation of aspects that are relevant for the hydrological regime such as seasonal flow, occurrence of extreme events, and monthly and daily pairwise indices (assess the skill to reproduce the observed time-series). The evaluation of these aspects allows identifying the capabilities and weaknesses of the impact assessments. Here, the simulations are evaluated against gauged data, working as a mean to assess the plausibility of the simulation outputs using uncorrected and bias-corrected RCMs. This work builds on studies that have assessed climate variables. For instance, the bias-corrected Euro-CORDEX simulations, at both resolutions, have a similar skill at capturing the wet-day intensity and precipitation frequency (Casanueva et al., 2016).

Here, we therefore address the two above-mentioned research aims by evaluating the simulation skill of five uncorrected and bias-corrected Euro-CORDEX RCMs at 0.11° and 0.44° using a range of temperature, precipitation and river flow indices, evaluating the mean along with high and low extremes, frequency of occurrence and daily and monthly simulation sequence. By using a multi-model ensemble, this analysis provides a robust understanding of the added value of high-resolution simulations and post-processing approaches for hydrological impact studies. We analyse four diverse catchments across Great Britain, representative of different climate and physical characteristics, focusing on the following questions:

1) Based on a range of selected indices, is the performance of the 0.11° Euro-CORDEX RCMs better than their 0.44° version to simulate (a) climate and (b) river flow?

2) Is the current skill of the Euro-CORDEX RCMs sufficient to generate plausible inputs for the analysis of climate change impacts on hydrology and how does this compare to the inputs from bias-corrected simulations?

3) Is there any improvement in the simulation skill of precipitation and river flow when using a Double Gamma Quantile Mapping (DGQM) bias correction compared to the usual Gamma Quantile Mapping (GQM) approach?

Given the associated computational cost (Bucchignani et al., 2015) and the potential for improving the skill of climate simulations, especially for impact assessments (Ehret et al., 2012), there is a clear need for rigorous evaluation of the added value of increasing RCM resolution. Previous hydrological impact studies have analysed this issue using one or two RCMs (e.g. Mendoza et al., 2016; Kay et al., 2015). However, their results might not be transferable to other RCMs, as each has its own parameterisations.

GQM inflates the precipitation extremes, producing unreliable flood simulations. Whilst this is a known issue (Cloke et al., 2013; Huang et al., 2014), no study has exhaustively compared the results between using the GQM and the DGQM approaches using extreme indices. This study provides a comprehensive analysis of such gaps.

## 2. Data and method

### 2.1. Observation databases and study catchments

We use the Centre for Ecology and Hydrology (CEH) Gridded Estimates of Areal Rainfall (CEH-GEAR) dataset as 1-km gridded daily precipitation observations (Keller et al., 2015). Records from the Natural Environment Research Council (NERC) Hydrology and Ecology Research Support System (CHESS) (Robinson et al., 2017) are used as 1-km gridded daily temperature observations. The 1-km gridded CHESS-PET dataset is employed as potential evapotranspiration (PET) observational reference. CHESS-PET uses the Penman-Monteith equation (Monteith, 1965) to calculate daily PET using climate variables from the Met Office Rainfall and Evaporation Calculation System (MORECS) (Hough and Jones, 1997) as input. We use river flow observations from the CEH's National River Flow Archive (NRFA).

We analyse four catchments within the UK. The catchments have long river flow records and cover regions that are representative of the different climate and catchment types that can be found within the UK. These are the Upper Thames, Glaslyn, Calder and Coquet catchments (Fig. 1). This set of catchments with different characteristics (Table 1) can aid identifying key features that impact on the simulation skill of the RCMs. The smallest catchment is the Glaslyn, which has the most complex topography and highest rainfall. The largest catchment is the Upper Thames (1616 $km^2$), which also has the least complex topography. The Calder and Coquet are intermediate in terms of area, elevation and precipitation. These catchments have been studied before using bias-corrected climate projections (QM, normal distribution for temperature and Gamma distribution for precipitation) from the HadRM3-PPE RCM (Prudhomme et al., 2013).

### 2.2. RCMs

We evaluate two spatial resolutions (0.11° equivalent to 12.5 km and 0.44° equivalent to 50 km) of five Euro-CORDEX RCMs (Jacob et al., 2014) driven by the ERA-Interim reanalysis (Dee et al., 2011), the so-called 'evaluation simulations'. The evaluation simulations are used as these are driven by observations and consequently simulate the internal variability in synchronicity with reality, in contrast to the historical simulations. The assessed RCMs are shown in Table 2 (refer to Table 1 in Kotlarski et al. (2014) and Table 1 in Prein et al. (2015) for a detailed RCM description). These models are selected as they have the best performance to reproduce observations in the British Isles according to Kotlarski et al. (2014). When more than one RCM cell is needed to fully cover the catchment we use the mean of the cells to represent the catchment's climate simulations (see Fig. 1).

### 2.3. Bias correction

QM is used based on parametric representations of the simulated and observed distributions (Piani et al., 2010). For each month of the year, the Gamma distribution is fitted to the daily precipitation and the normal distribution to the daily temperature. RCMs generally simulate too many days with very low precipitation and not enough dry days. Therefore, in an initial step the QM method adjusts the number of simulated dry days such that they match with the number observed dry days by including a wet day threshold and replacing all values below it with zero. After the wet-day adjustment, the distributions of the simulations and observations are matched using their cumulative distribution functions (CDF). The method is represented by the following equations:

$$P_c(t) = F_g^{-1}\big(F_g(P_R(t), \alpha_R, \beta_R), \alpha_O, \beta_O\big) \tag{1}$$

$$T_c(t) = F_n^{-1}(F_n(T_R(t), \mu_R, \sigma_R^2), \mu_O, \sigma_O^2) \tag{2}$$

Where $P_c(t)$ and $P_R(t)$ represent the bias-corrected and raw RCM daily precipitation, respectively. Likewise, $T_c(t)$ and $T_R(t)$ stand for the bias-corrected and raw RCM daily temperature. The raw RCM CDF is symbolized with F, and $F^{-1}$ stands for the observations inverse CDF. The 'g' and 'n' subscripts represent the Gamma and normal distributions, respectively. The precipitation shape and scale parameters are symbolised by α and β and the temperature mean and standard deviation by μ and σ, respectively. Finally, the 'R' and 'O' subscripts are used to symbolize the distribution parameters from the raw RCM and observations, respectively.

GQM focuses on the most frequent values (e.g. means) (Teng et al., 2015; Yang et al., 2010). Consequently, the corrected precipitation extremes tend to be inflated compared to the observations (Cannon et al., 2015). Therefore, we also bias-correct precipitation using the DGQM. The methodology is mainly the same as the GQM with the difference that the simulated precipitation distribution is divided in two segments. Each is corrected separately, generating correction parameters for each section. In our study, the distribution is divided at the 90th percentile because at this percentile the biases inflate (see section 3.2.2.1).

For the 0.11° RCMs, the spatial scale of the simulations and the observations are approximately the same and the method can be viewed as a pure bias correction. In contrast, the output of the 0.44° is given on a larger scale than the observations and thus the QM also includes a downscaling aspect to account for the difference in distributions on different spatial scales. We note that due to the existence of sub-grid variability QM is in principle problematic as the corrected values for all sub-grid locations would have unrealistic high correlations (Maraun, 2013). However, this limitation is not of high relevance for our study as we bias-correct the distributions for the entire catchments.

### 2.4. Hydrologic simulation

The Hydrological Modeling System from the US Army Hydrologic Engineering Center (HEC-HMS) (Scharffenberg, 2013) is used to simulate the catchments' daily river flow. HEC-HMS has been successfully used before to analyse climate change impacts on water resources in other regions (e.g. Babel et al., 2014; Azmat et al., 2015). An advantage of the model is the available guidance for the estimation of parameters. Here, the model is run using its continuous, lumped arrangement. Observed precipitation and PET time series are used as input for the calibration and validation of the model. Afterwards, the raw and bias-corrected RCM simulations drive the model to generate the river flow simulations.

Evapotranspiration controls the moisture returning from the Earth's surface to the atmosphere and therefore impacts on the river flow. PET estimates the amount of water returning to the atmosphere when enough water is present in the surface of the catchment. Climate models do not simulate PET directly, thus it is estimated indirectly with formulas using variables from the climate models as input. There is no consensus on whether temperature–based or physically-based formulas provide better results in a climate change context (Kay et al., 2013) as the data required by the physically-based formulas is uncertain in the climate model simulations compared to the input from one variable formulas (Kingston et al., 2009). This has been discussed and explored elsewhere (please refer to: Seiller and Anctil, 2016; Kingston et al., 2009; Kay and Davies, 2008; Kay et al., 2013). We estimate PET using the Oudin formula (Oudin et al., 2005) as it has given accurate results before (e.g. Oudin et al., 2005; Kay and Davies, 2008).

$$\begin{cases} PET \ (mm \ day^{-1}) = \frac{R_e}{\lambda \rho} \left( \frac{T+5}{100} \right) & if \ T + 5 > 0 \\ PET \ (mm \ day^{-1}) = 0 & otherwise \end{cases} \qquad (3)$$

The extraterrestrial solar radiation ($R_e$) is the solar radiation received at the top of the Earth's atmosphere which can be estimated by the latitude and day of the year. The density of water is symbolized by $\rho$, the latent heat flux by $\lambda$ (2.45 MJ/kg) (Allen et al., 1998) and T is the daily mean temperature (°C). When driven by observed temperature, the Oudin formula gave results similar to the CHESS-PET dataset for 1973 to 2010 (Pasten-Zapata, 2017).

### 2.5. Hydrological model calibration

The hydrological model is calibrated and validated against the observations using a split sample test. Considering the available uninterrupted daily river flow records, for each catchment two same-length non-overlapping time periods are used: one for calibration and the other for validation. Three indices are assessed: the low flows simulation is evaluated using the Q95 (flow equalled or exceeded 95% of the time), the high flows by the Q10 (flow equalled or exceeded 10% of the time) and the Nash-Sutcliffe Efficiency Index (NSE) which evaluates the fit of the simulated and observed river flow. The NSE ranges from 1 (perfect fit) to negative (unreliable model) (Nash and Sutcliffe, 1970). In the NSE formula, $Q_t^{obs}$ and $Q_t^{sim}$ stand for the observed and simulated river flow at time step t, respectively. $Q^{mean}$ is the average of the observed river flows during the complete period.

$$NSE = 1 - \left[ \frac{\sum_{t=1}^{n} \left( Q_t^{obs} - Q_t^{sim} \right)^2}{\sum_{t=1}^{n} \left( Q_t^{obs} - Q^{mean} \right)^2} \right] \qquad (4)$$

### 2.6. RCM validation approach and indices

Validation is important to assess the RCM simulation skill before and after bias correction (Eden et al., 2014). Here, a five-fold cross-validation approach is used: 1) the study period is divided into five same-length, non-overlapping blocks, and 2) the QM methods are trained using four blocks and the remaining block is corrected using the parameters from the training period (Maraun et al., 2015). The corrected blocks are concatenated to time series for the entire period from which the performance measures for the bias-corrected precipitation and temperature are derived.

A range of distribution-based and time series-based indices evaluate the skill of the raw and bias-corrected RCM outputs to simulate precipitation, temperature and river flow. The indices assess biases in the means, low and high extremes, inter and intra annual variability and correlations for each variable (see Table 3). RCMs are then ranked based on their skill to simulate all indices relative to the skill of the other RCMs at both resolutions. We use the complete time series (dry days included) to estimate the precipitation indices. Even when driven by "perfect boundary conditions", a close similarity between the RCM simulations and observations is not expected (Kay et al., 2015) due to subgrid variability or internal variability because the boundary conditions do not fully determine the weather states within the RCM. Nevertheless, we include daily and monthly pairwise indices as these are important for the river flow simulation. We left out the hydrological model uncertainty source intentionally to solely evaluate the effects of increasing RCM resolution. Thus, we compare the river flow simulations driven by RCM outputs against the river flow simulations driven by the observed temperature and precipitation.

## 3. Results

This section begins by showing hydrological model simulation skill followed by the evaluation of the simulation skill of the uncorrected RCMs for temperature, precipitation and river flow. Finally, we compare the biases that remain after bias-correcting precipitation using the GQM and DGQM and their impacts on the river flow simulation.

### 3.1. Calibration and validation of the hydrological model

Firstly we evaluate the hydrological model simulation skill using climate observations as input. Depending on the catchment, the length of the overall evaluation period ranges from 34 to 49 years. The daily NSE varies between 0.62 (Calder) and 0.78 (Glaslyn) for calibration and between 0.52 (Coquet) to 0.78 (Glaslyn) for validation (Table 4). These results indicate a moderate to good simulation skill overall compared to the NSE values from similar studies which vary from 0.45 to 0.9 (e.g. Arnell, 2011; Cloke et al., 2013). The Q10 bias ranges between -6% and 11% for the calibration and between –5% and 7% for the validation. Similarly, the Q95 bias ranges between -27% and -11% for the calibration and between -44% and 6% for the validation. Overall, the simulation of high flows is very good and moderate to very good for the low flows.

### 3.2. Evaluation of the RCM simulation skill

We now assess the skill of the RCMs at simulating climate and river flow, firstly for the raw simulations and then for the bias-corrected outputs. We only show robust results for the analysis of the indices (e.g. if all RCMs from a particular resolution underestimate or overestimate an index). We also evaluate the multi-model percentile bias for each variable and use a skill rank to enable comparison of the different RCMs over the different performance indices. The ranking is only estimated for the uncorrected simulations as the biases after the correction are small and similar among the RCMs. Thus, ranking the bias-corrected simulations would give meaningless results.

#### 3.2.1. Uncorrected RCM simulations

##### 3.2.1.1. Temperature

We begin with assessing the ability of the RCMs to simulate temperature. The 0.11° RCMs underestimate the annual mean temperature for the upper Thames (Fig. 2a, ii), Calder (Fig. 2c, ii) and the Coquet (Fig. 2d, ii) catchments, whereas the 0.44° RCMs overestimate the annual mean temperature for the Glaslyn (Fig. 2b, ii) and Coquet (Fig. 2d, ii) catchments. The monthly mean temperature bias for the 0.11° RCMs is larger for the Calder (between and 0.5 °C and 1.1 °C) (Fig. 2c, ii) and smaller for the Glaslyn catchment (between 0.4 °C and 0.7 °C) (Fig. 2b, ii). In contrast, the monthly mean temperature bias of the 0.44° RCMs is larger for the Glaslyn (between 0.4 °C and 1.2 °C) (Fig. 2b, ii) and smaller for the Calder catchment (between 0.8 °C and 1.0 °C) (Fig. 2c, ii).

We use the simulation spread to evaluate the simulation skill of each resolution. The spread represents the range between the highest and lowest simulated value considering all RCMs at each resolution and all gridcells within a catchment. The temperature percentile bias spread for the upper Thames is similar for both resolutions except between the 40th and 60th percentile where the 0.44° simulation include larger positive biases (Fig. 3a). For the Glaslyn catchment, the 0.44° simulations overestimate temperature for almost all percentiles, while the biases of the 0.11° simulations are smaller (Fig. 3b). For the Calder catchment, the 0.44° RCM spread includes the no bias threshold for all percentiles, whereas the 0.11° RCMs underestimate temperature between the 40th and 90th percentile (Fig. 3c). Finally, in the Coquet catchment the 0.44° simulations overestimate temperature below the 70th percentile and the 0.11° simulations underestimate it between the 40th and 80th percentiles (Fig. 3d). The Pearson correlation coefficients of the daily time series vary between 0.91 and 0.97 in all catchments for both resolutions (Figs. 2, iii).

Integrating the RCM simulation skill of all the indices into a ranking shows that, in the upper Thames, two out of five high-resolution uncorrected simulations outperform their 0.44° version (last column of Table 5). Similarly, for the Calder catchment, one 0.11° simulation outperforms its 0.44° version and all five high-resolution simulations outperform their low-resolution version for the Glaslyn and Coquet catchments. This indicates that topography has an influence in the simulation of temperature and RCM resolution has an effect in the simulation skill for catchments with larger elevation variability where, for observations at high elevation, the 0.44° RCMs would be expected to have positive biases as the grid elevation is lower that the observations.

Based on the rank, the overall best performing simulation for the upper Thames and Calder catchments is 0.44° RACMO, whereas for the Glaslyn and Coquet catchments, the 0.11° RACMO and HIRHAM simulations, respectively, outperform the rest. This implies that biases from the high-resolution simulations are smaller for the catchments with complex topography, which is better represented by the 0.11° simulations. The biases are a consequence of systematic model biases in the elevation and a lack of representation of the elevation variability. Nevertheless, for larger and flatter catchments the simulation skill from both resolutions is similar.

### 3.2.1.2. Precipitation

Now we assess the skill of the uncorrected RCMs to simulate precipitation. Overall, RCMs have biases when simulating extremes. For instance, the SDII ratio is underestimated in all catchments by the 0.44° simulations (Figs. 4a, S1a and S2a), except for the Coquet (Fig. S3a). In all catchments the RX1day is overestimated by both resolutions between 24% and 93%. The R10 and R20 are underestimated at the Glaslyn catchment between -23 and -77 days and between -16 and -45 days, respectively (Fig. S1d). Similarly, in the Calder catchment R10 and R20 are underestimated by the 0.44° simulations between -5 and -10 days and between -3 and -4 days, respectively (Fig. S2d). These results indicate that the uncorrected models can provide unrealistic simulations of extreme precipitation.

It is expected that the models simulate the precipitation mean better than the extremes. Even though the spread of the models includes the observed mean precipitation for most catchments, there are cases when this does not happen. The annual mean precipitation is underestimated by both resolutions in the Glaslyn catchment between -22% and -67% (Fig. S1c). This may be because the analysed RCMs do not correctly simulate convective precipitation. In the Calder catchment the 0.44° simulations underestimate the annual mean precipitation between -7% and -16% (Fig. S2c). This can be due to local precipitation not being correctly simulated by the coarse models. The absolute monthly mean precipitation bias for both resolutions varies between 7% and 67% in all study cases (Figs. 4c, S1c, S2c and S3c).

The simulated precipitation bias spread increases in all catchments as the percentile increases. The spread of the 0.11° simulations is larger than for the 0.44° simulations (Fig. 5, first row). In the upper Thames catchment, the 0.11° simulations reach their largest spread, -1 to 4 mm/day, above the 95th percentile whereas the largest spread of the 0.44° RCMs ranges from -1 to 1 mm/day (Fig. 5a). In the Glaslyn catchment, the bias spread deviates from the observations at the 50th percentile for the 0.44° simulations and at the 60th percentile for the 0.11° simulations (Fig. 5d). In the Calder catchment, the 0.11° simulations spread includes the no bias threshold for the whole distribution whereas the 0.44° simulations spread deviates from that threshold at the 70th percentile (Fig. 5g). In the Coquet catchment, the spread from both resolutions includes the zero bias threshold for almost all percentiles (Fig. 5j).

The dry and wet spell biases are important for the simulation of river flow as this is influenced by the daily sequence of the wet/dry conditions. The absolute dry spell bias for both resolutions in all catchments range between 0.2 to 1.6 days, with a similar simulation skill in all catchments (Figs. 4b, S1b, S2b, S3b). Likewise, the absolute wet spell bias for both resolutions varies between 0.1 and 1.6 days in all catchments (Figs. 4b, S1b, S2b, S3b). Biases in the upper Thames for this measure are smaller, 0.2 to 0.6 days (Fig. 4b), compared to the other catchments. These results do not show large simulation biases. Considering the time-series based indices, correlation coefficients are above 0.4 and below 0.8 in all catchments, showing differences between the daily observations and simulations (Figs. 4c, S1c, S2c, S3c).

Considering the ranking for all indices, only for the Glaslyn catchment do all the 0.11° simulations outperform their 0.44° version (Table 6, last column). From the five RCMs, two 0.11° simulations outperform their low-resolution version for the Upper Thames and three for the Calder and Coquet catchments. The 0.11° CCLM and WRF have better simulation skill than their 0.44° version in all catchments. In contrast, for HIRHAM and RCA, the improvement is only observed in one catchment. For the latter models, there is no added value from increasing the resolution as the simulation processes occurring at higher resolutions than the 0.44° gridbox do not improve the results, possibly due to an inappropriate physical representation. The 0.11° CCLM is the best performer in all catchments, except for the Glaslyn where 0.11° HIRHAM has the highest rank.

All high-resolution simulations outperform their coarse simulations at the Glaslyn catchment due to the differences between the sizes of the catchment and the different cells. Thus, increasing the RCM resolution increases their simulation skill for catchments with larger elevation variability because the RCMs are able to represent the high-resolution features. In general, increasing the RCM resolution reduces the simulation biases in the upper tail of the distribution, but there are also high-resolution models that consistently overestimate precipitation (e.g. RCA in Figs. 4, S1, S2, S3). The low-resolution models do not simulate the small-sized catchment accurately. In contrast, the flat and large catchments are simulated similarly by both resolutions, showing no added value from increasing RCM resolution.

### 3.2.1.3.    River flow

Now, we evaluate the RCM skill in providing inputs for simulating the river flow in each catchment. In the upper Thames, the 0.11° RCMs overestimate the spring discharge by between 16% and 194% (Fig. 6a). Both resolutions underestimate all indices in the Glaslyn catchment (Fig, S4). In the Calder catchment, the 0.44° RCMs underestimate the annual (-9% to -31%) and autumn (-10% to -50%) flows, whereas the 0.11° RCMs overestimate the discharge during winter (3% to 63%) and spring (22% to 104%) (Fig. S5a). Also, the Q10 and Q10 annual frequency are underestimated by the 0.44° RCMs (Fig. S5b and c). In the Coquet catchment, the winter mean discharge is underestimated by the 0.44° RCMs by between -7% and -42% and during summer it is overestimated by the 0.11° RCMs by between 2% and 218% (Fig. S6a). In addition, the Q95 is overestimated by the 0.11° simulations.

Except for the Glaslyn catchment, the multi-model simulation spread of the flow duration curve (FDC) from both resolutions includes the observed FDC entirely (Fig. 7, first row). For the Glaslyn catchment, both resolutions underestimate the FDC with the 0.11° simulation spread being closer to the observed FDC (Fig. 7d). The 0.44° simulation spread is larger than the 0.11° spread in the Coquet, but smaller in the upper Thames. In the remaining catchments, the spreads of both resolutions are similar.

Overall, the maximum monthly NSE values are 0.42 for the Upper Thames (Fig. 6e), 0.22 for the Glaslyn (Fig. S4e), 0.67 for the Calder (Fig. S5e) 0.26 for the Coquet catchment (Fig. S6e), indicating that the best river flow simulation is moderate to poor for all catchments except for the Calder.  In contrast, the minimum NSE values are negative in all catchments, implying that there are RCM outputs that generate unreliable river flow simulations even at the monthly times step. Negative NSE values can be a result of river flow overestimation in all indices, for instance 0.11° RCA and HIRHAM in the Calder and Coquet catchments. The Spearman correlation coefficients of the daily river flow are higher for the upper Thames and Calder and smaller for the Glaslyn and Coquet, indicating that the RCMs are able to simulate the daily river flow sequence better on the large and flat sites compared to the small and topographically-complex catchments (Fig. 67f, S4f, S5f and S6f).

Comparing their skill in simulating all indices by means of their rank, three 0.11° simulations outperform their 0.44° version in the Upper Thames, five in the Glaslyn, one in the Calder and two in the Coquet catchment (Table 7, last column). Overall, for both resolutions, biases in particular indices are large and the skill of the pairwise indices (NSE, MSE, correlation) is low. The 0.11° simulation biases are consistently smaller than the 0.44° biases only for the Glaslyn catchment due to the difference between the catchment and the 0.44° RCM cell size. However, for this catchment biases are large even for the high-resolution simulations indicating that subgrid processes that result in precipitation increases are not represented by the models. Only CCLM gives better simulation skill for its high-resolution in all catchments.

### 3.2.2. Bias-corrected RCM simulations

#### 3.2.2.1.    Temperature

Bias-correction reduces the mean and percentile biases by construction (Figs. 3e,f,g,h). Thus, the skill of all RCMs becomes similar in all catchments, as expected. Overall, the larger distribution biases are for the 1st and 99th temperature percentiles, with biases lower than 1°C (Figs. 2, i). Even though these percentiles have the largest biases after bias correction, as may be expected the biases are smaller than those of the uncorrected RCMs. QM does not improve the daily sequence simulation. As a consequence, there is only a slight change in the Pearson correlation coefficient of the daily time series (Figs. 2, iii).

### 3.2.2.2. Precipitation

#### 3.2.2.2.1. Gamma distribution QM

The skill of both RCM resolutions becomes similar after application of GQM. Nevertheless, biases are not reduced for the 95th percentile, SDII ratio, wet spell length, R95p and R20 in the Upper Thames, for RX1day in the Calder and for the SDII ratio in the Coquet catchment. These indices evaluate the extremes, which are inflated by the correction method (Cannon et al., 2015), and the precipitation intensity.

Considering the indices that are not based on the distribution, the Spearman correlation slightly increases after GQM (Figs. 4c, S1c, S2c and S3c) whereas for the MSE the multi-model ensemble bias is reduced, but there are cases when the biases of individual RCMs increase (Figs. 4c, S1c, S2c and S3c). The same happens for the wet and dry spell lengths (Figs. 4b, S1b, S2b and S3b) and RX1day (Figs. 4c, S1c, S2c and S3c). The multi-model bias spread from both resolutions is similar and smaller than 1 mm/day up to the 90th percentile in all catchments (Fig. 5, second row). Above the 90th percentile, the spread of both resolutions increases exponentially. The bias spread in the extremes is larger for the Glaslyn catchment possibly as a consequence of the bias magnitude of the original uncorrected simulation (Fig. 5e).

#### 3.2.2.2.2. Double Gamma distribution QM

After applying the DGQM method, the skill with respect to distribution-based indices from all RCMs at both resolutions becomes similar. The biases for most distribution-based indices are reduced compared to both uncorrected and GQM. In all catchments, the biases are lower than 1 mm/day below the 99th percentile after which biases increase. Thus, DGQM reduces the percentile biases in all catchments compared to GQM. For the 90th precipitation percentile the DGQM approach increases the biases in all catchments because at this percentile the method segments the precipitation distribution, generating an increment in the bias. Nevertheless, this increase is approximately ± 1 mm/ day in all catchments except the Glaslyn. Additionally, the simulation bias spread of both resolutions is similar for all catchments, as expected (Fig. 5, last row).

For the extreme and precipitation intensity measures, DGQM reduces the biases compared to GQM except for the RX1day and SDII ratio in the Upper Thames , R20 in the Glaslyn, R10 in the Calder and the SDII ratio in the Coquet catchment. The simulation skill of the uncorrected models and the GQM and DGQM approaches is similar in all catchments for the Spearman daily correlation coefficient. Overall, the DGQM provides outputs with smaller biases for most of the indices compared to the uncorrected and GQM simulations.

### 3.2.2.3.    River flow

#### 3.2.2.3.1.    Gamma distribution QM

River flow is simulated using the GQM precipitation and temperature as drivers. GQM decreases the bias of all indices in every catchment, except for the Q10 in the upper Thames catchment (Fig. 6c). The bias-corrected FDC simulation spread decreases for both resolutions in all catchments (Fig. 7, second row). The observed FDC is completely included within the spread of both resolutions showing a good simulation of the entire distribution.

From the pairwise indices, the skill of the multi-model ensemble improves for the monthly NSE (Fig. 6e) and the spread of the daily MSE is reduced in most cases. However, GQM can result in negative NSE values for some models that had positive values when these were not bias-corrected (e.g. 0.44° RACMO and HIRHAM in the Upper Thames). The Spearman correlation of daily time series increases slightly in all cases (Fig. 6f, S4f, S5f and S6i).

#### 3.2.2.3.2.    Double Gamma distribution QM

The DGQM approach decreases the biases for all the distribution-based indices compared to both uncorrected and GQM with the exception of Q95 for the Glaslyn catchment. Considering the non-distribution-based indices, the NSE and MSE are not improved for the Coquet catchment. Even though the biases are reduced, the simulation skill among all RCMs does not become similar for specific cases with indices involving the extremes and the pairwise simulation (e.g. the Q10 annual frequency, Q10 and NSE for the Upper Thames, Fig. 6b,c and e). Overall, the daily MSE and monthly NSE simulation skill improves compared to the GQM approach. Thus implying that the river flow simulation skill is better when using the DGQM. By construction of the bias correction method, the FDC simulation spread of both resolutions is similar in shape and amplitude (Fig. 7, bottom row). Compared to GQM, the DGQM simulation spread is further reduced.

The Spearman correlation coefficient of the daily river flow time series increases slightly with not a large difference compared to the GQM simulations. Overall, applying the DGQM approach results in smaller biases compared to the GQM, in specific for the simulations of extremes and the monthly sequence.

## 4. Discussion

Regarding our first research question, as to whether the relative performance of the high- resolution simulations is better than that of the lower-resolution simulations, the results show that the high-resolution RCMs consistently have a better simulation skill for climate and river flow only in the Glaslyn catchment. This is mainly because the size of this catchment is smaller than the 0.44° RCM cell, and it has a complex topography and high precipitation. As a consequence, the skill of the 0.44° simulations in reproducing the local physical features of this catchment is not good. For the other catchments, all of which are larger in size and with less complex topography and less precipitation, both resolutions have a similar performance. Similar results were obtained for the Upper Danube using HIRHAM at resolutions of 50km and 12km (Dankers et al., 2007). Only the skill of CCLM improved when using the high-resolution version. Kotlarsky et al. (2014) found that CCLM also gave good results when simulating the mean, seasonal and 95[th] percentile of precipitation over the British Isles. In our study, the remaining RCMs did not improve their simulation skill, implying that the high-resolution versions of these models do not accurately represent processes occurring at higher resolutions.

The performance of the two RCM resolutions at simulating temperature was clearly linked to the topographic characteristics of the study catchments. In the upper Thames and Calder catchments, which have relatively flat topography, we found that there is no clear added value from the uncorrected high-resolution RCMs; however, in the topographically-complex Glaslyn and Coquet catchments, all 0.11° simulations outperformed their 0.44° version. These findings are similar to that of Onol et al. (2012) and Tolika et al. (2016) and it is likely that they can be attributed to the difference in elevation from the grid cells of the observations and models, and the lack of representation of the spatial variability. Increases in the simulation skill of local climate when using higher-resolution simulations have been reported before, particularly for mountainous regions (Evans et al., 2013; Larsen et al., 2013; Tolika et al., 2016).

The uncorrected 0.11° simulations largely underestimate the precipitation and river flow observations of the Glaslyn catchment, mainly due to the catchment's topographic complexity and high levels of precipitation. Similar results for the Euro-CORDEX RCMs have been obtained for precipitation in other regions with complex topography (e.g. Casanueva et al., 2016; Prein et al., 2015; Torma et al., 2015). For the remaining catchments, the multi-model simulation spread of the simulations of both resolutions includes the observed FDC, indicating that the models are able to provide useful simulations that resemble the observed river flow. However, the simulation spread can be large; deviations in the annual mean river flow reach almost 100% for some RCMs. Individual uncorrected RCMs have small biases and satisfactory simulations of the river flow (e.g. 0.11° CCLM in the Calder and Coquet catchments), but there are also RCMs that are not able to provide useful simulations. For example, the 0.11° RCA had the largest precipitation and river flow biases in most indices for all catchments. In contrast, all the bias-corrected RCM simulations are closer to the observed climate and river flow. Furthermore, the simulation skill from all bias-corrected RCMs at both resolutions becomes similar and as a result, the simulation spread of the multi-model ensemble is reduced compared to the uncorrected simulations, providing a smaller range of possible scenarios.

Our results show that uncorrected RCMs provide river flow simulations that have too much spread to be able to be used for impact studies (also stated by Kay et al., 2015; Cloke et al., 2013). Both resolutions have a similar performance when simulating the seasonal mean river flow as there are biases from both resolutions. However, certain high-resolution models tend to overestimate the seasonal flow largely for most of the catchments and seasons (e.g. RCA in all catchments and HIRHAM in the Coquet and Calder catchments). In contrast, the low-resolution CCLM underestimates river flow for all seasons and catchments. For the medium-sized Calder catchment, individual models have different biases per season but the multi-model ensemble mean shows a consistent underestimation for high-resolution models and underestimation of river flow for the low-resolution modes. This is not distinguished in the larger Upper Thames nor in the Coquet catchment. Similar to the annual mean flow, both resolutions underestimate the seasonal flow in the Glaslyn catchment. In comparison, all the bias-corrected RCMs simulate the river flow much closer to the observed flows and reduce the simulation spread, thus providing plausible inputs for impact studies.

Finally, to answer our last research question, we evaluate the simulation skill of DGQM compared to GQM. Using four catchments with different characteristics, the DGQM provides a better simulation of the river flow characteristics compared to the QGM approach, with a higher improvement for the simulation of extremes and the monthly sequence. The GQM systematically reduces the precipitation bias up to the 90th percentile, but exponentially increases the bias above this percentile. Therefore, to capture the properties of extremes, we suggest using the DGQM with the 90th percentile as segmentation threshold in contrast to Yang et al. (2010) who divided the distribution at the 95th percentile. Based on our results, the DGQM reduces the precipitation and river flow biases of most indices compared to the commonly used GQM. This is particularly relevant for the analysis of extreme precipitation and high flows as the GQM is usually employed in flood analysis (e.g. Cloke et al., 2013) and river flow projections (e.g. Prudhomme et al., 2013). In addition, the DGQM reduces the ensemble spread more than the GQM, without introducing much extra complexity. However, no bias correction method will remove all biases. Thus, the selection of the method depends on the requirements of each study (Nguyen et al., 2017) and it should be tested to evaluate whether the benefits justify their calculation complexities.

Ideally, RCMs should not require post-processing techniques to provide simulations which can be used with confidence (Ehret et al., 2012). However, our results demonstrate large biases for various diagnostic indices for the reanalysis-driven RCMs. Particular RCMs provide plausible river flow simulations, for instance, 0.11° CCLM for the Calder catchment when assessing the annual and seasonal means, low flows, high flow occurrence and pairwise indices. However, the RCM simulation skill is catchment-dependent. Thus, at the moment, bias correction seems to be the best approach to reduce the ensemble spread and its biases. Nevertheless, bias correction methods should be used carefully for the analysis of future projections (Cloke et al., 2013) as bias correction cannot correct fundamental problems from the original climate model (Maraun and Widmann, 2015; Maraun et al., 2017) and the spread of the bias-corrected simulations might not reflect the total real uncertainty. Climate research is focusing on determining the causes behind the biases (e.g. Addor et al., 2016) and improving the simulation of the processes (e.g. Zittis et al., 2017; Meredith et al., 2015). For instance, convection permitting models seek to improve the simulation of precipitation extremes (Tölle et al., 2017; Gutjahr et al., 2016). However, the computational cost of developing such models is large and, as a consequence, the simulation length is short and the availability of GCM-RCM projections is low.

By analysing four catchments with different characteristics, we evaluate the RCM simulation skill in different contexts. Our results suggest that the small size and the high precipitation (e.g. Glaslyn catchment) are the main factors related to the better simulation skill from the high-resolution RCMs over the low-resolution models for the simulation of river flow. The importance of topographical complexity and other characteristics for the simulation outputs is secondary. This is highlighted by the results of the medium-sized Coquet catchment, for which both resolutions have similar simulation skill even with its complex topography.

## 5. Conclusions

This study provides information on the added value from increasing RCM resolution and bias correction techniques for the simulation for river flow. Previous studies have assessed the improvement in the simulation skill of climate variables due to an increase in the RCM resolution, but this might not guarantee an improvement in the simulation of the river flow parameters that are relevant for impact studies. We conducted a comprehensive analysis on how the uncorrected and bias-corrected RCM outputs drive the simulations of river flow at high and low resolutions. Each RCM used here has the same parameterization, domain and driving data at both resolutions, and therefore the comparison only evaluates the effect of increasing its resolution. We analysed four catchments located at different latitudes within Great Britain. These catchments vary in climate (e.g. precipitation ranging from 2900 mm $yr^{-1}$ to 762 mm $yr^{-1}$), physical characteristics (flat and complex topographies, areas ranging from 69 $km^2$ to 1616 $km^2$), land use (varying from urban-dominant to agricultural and natural areas) and hydrological characteristics (e.g. annual mean river flow ranging from 15.3 $m^3$ $s^{-1}$ to 5.8 $m^3$ $s^{-1}$). We applied a detailed assessment of the simulation skill of the climate and hydrological models using a set of indices relevant for the analysis of different impacts.

We show that the uncorrected 0.11° RCMs only showed better skill in simulating precipitation and river flow in the small catchment. This is because the spatial resolution of the 0.44° models is four-times larger than the catchment size, whereas one cell of the 0.11° model is similar in area to the catchment. Nevertheless, the high-resolution simulations are not able to accurately represent the complex topography of this catchment and do not resolve local processes, underestimating the observed precipitation and the entire FDC.

Both resolutions capture the temperature and precipitation distribution, as well as the FDC, for the remaining sites. Thus, in principle, the simulations could be used for climate change assessments. Nevertheless, for most of the indices, the multi-model variability is large (e.g. the mpe of the annual mean river flow simulation ranges from 198% to -31% in the Upper Thames, with an average of 49%), making any interpretation difficult in practice. Only one RCM (CCLM) improves the river flow simulation when using its high-resolution version in all catchments, implying that the remaining models do not simulate the relevant high-resolution processes accurately as there is no consistent difference between their high and low resolution versions. Therefore, there is no added value from using the high-resolution RCMs in those catchments for the assessment of river flow impacts.

Bias-correction reduces the distribution-based biases for all RCMs and resolutions by construction. Thus, the bias-corrected high- and low-resolution RCMs have similar simulation skill for the distribution-based indices. There is also less spread from the ensemble simulation of precipitation and river flows (e.g. the mpe of the annual mean river flow simulations for the Upper Thames ranges from -1% to 16% when corrected using DGQM, with an average of 7%). Nevertheless, daily pairwise indices, which assess the skill of the model when simulating the observed time series, are not improved by bias correction. However, the monthly NSE results indicate that bias correction can improve the pairwise simulation on monthly timescales. Overall, correcting the RCMs to the local temperature and precipitation provides a reduction of the ensemble spread, making the outputs more useful for the analysis of impacts. Nevertheless, it should be considered that the ensemble spread of uncorrected and corrected models can underestimate the true simulation uncertainty.

In comparison to GQM, DGQM provides a larger reduction in the simulation biases for precipitation and river flow. The main difference between both methods is the greater correction from DGQM for precipitation extremes (95th percentile, R10, R20, R95p) and high flows (Q10 and Q10 annual frequency). The monthly NSE consistently shows an improvement in the simulation skill of RCMs that are corrected using DGQM. Overall, for most of the RCMs and considering the results from all indices, the DGWM outperforms GQM.

Our study shows that an increase in RCM resolution does not always imply a better simulation of hydrological impacts, especially for large catchments. In contrast, small catchments with complex topography are still difficult to be simulated accurately by high-resolution models, concurring with Dankers et al (2007). The uncorrected RCM ensemble generally provides a large spread which makes it difficult to use for impact assessment. Bias-correction provides an alternative to reduce the biases and multi-model spread, making decision-making easier. From the methods evaluated here, DGQM reduces most of the RCM biases without much more complexity added to the bias-correction method employed when using GQM. However, and agreeing with Cloke et al. (2013) and Huang et al. (2014), the bias-corrected outputs should be used carefully when evaluating changes in very extreme flows as the correction inflates the simulated extremes. Compared to previous studies, we can state that our results are robust as we included a larger number of RCMs with different parameterizations for our analysis.

Whilst effective, bias correction adds extra uncertainty to the analysis chain (Cloke et al., 2013; Rummukainen, 2016). Therefore, it must be used with consideration of its limitations: dependence on the training period (Lafon et al., 2013), assumption of temporal stability of the correction function (Chen et al., 2015), issues of sub-grid variability and inflation of variance (Maraun, 2013), inter-variable consistency (Wilcke et al., 2013), spatial representation over complex terrain (Maraun and Widmann, 2015) and biases from the driving data (Maraun et al., 2017). The extent to which the climate change signal is altered must also be considered (Maraun, 2013; Velázquez et al., 2015) along with the possibility that bias correction can produce larger biases for extremes than for the mean (Huang et al., 2014). Additionally, we acknowledge that using different data to drive the RCMs used in this study, for instance a GCM, could give different results.

Our results are potentially useful for regions where RCMs of high(er) resolution are not yet developed and impact users require from solid knowledge to determine whether the high-resolution simulations are needed or not. These regions can gain knowledge from the added value of increasing RCM resolution that is evaluated here using a large set of indices. If used, bias-correction methods should be tested for the specific analysis that will be performed. This study provided different methods to perform this testing for the different RCMs and bias-correction methods for climatology and hydrology.

**Acknowledgements**

Pastén-Zapata received funding from Consejo Nacional de Ciencia y Tecnología (Conacyt) and Secretaría de Educación Pública (SEP) during the development of this study.

**Tables**

Table 1. Characteristics of the study sites

| | Upper Thames | Glaslyn | Calder | Coquet |
|---|---|---|---|---|
| Area (km$^2$) | 1616 | 69 | 316 | 346 |
| Maximum elevation (masl[1]) | 330 | 1080 | 556 | 775 |
| Minimum elevation (masl[1]) | 52 | 30 | 40 | 71 |
| Mean annual precipitation (mm/year) | 762 | 2957 | 1251 | 968 |
| Mean annual temperature (°C) | 9.7 | 8.1 | 8.4 | 7.4 |
| Mean annual PET (mm/yr) | 522 | 477 | 486 | 473 |
| Mean annual river flow (m$^3$/s) | 15.3 | 5.8 | 8.8 | 6.1 |
| Precipitation 90th percentile (mm/day) | 6.7 | 24.4 | 10.3 | 7.7 |
| Precipitation 95th percentile (mm/day) | 10.2 | 34.2 | 14.8 | 11.9 |
| [2]Q10 (m$^3$/s) | 34.8 | 13.5 | 19.9 | 12.4 |
| [3]Q95 (m$^3$/s) | 1.90 | 0.55 | 1.99 | 0.84 |

[1] Meters above the sea level
[2] River flow that is exceeded for 10% of the daily river flow time series
[3] River flow that is exceeded for 95% of the daily river flow time series

Table 2. RCMs used in this study

| RCM | Institute | Period | Reference |
|---|---|---|---|
| CCLM-CLMCOM | Brandenburg University of Technology (BTU) | 1989-2008 | Böhm et al., 2006; Rockel et al., 2008 |
| HIRHAM 5 | Danish Meteorological Institute (DMI) | 1989-2008 | Christensen et al., 1998 |
| RACMO22E | Royal Netherlands Meteorological Institute (KNMI) | 1979-2008 | Van Meijgaard et al., 2012 |
| RCA4 | Swedish Meteorological and Hydrological Institute (SMHI) | 1984-2008 | Samuelsson et al., 2011 |
| WRF 3.3.1 | Institute Pierre Simon Laplace (IPSL) and Institute National de l'Environment Industriel et des Risques (INERIS) | 1989-2008 | Skamarock et al., 2008 |

Table 3. Description of the precipitation, temperature and river flow indices used in this study

| Index | Description | Performance measure |
|---|---|---|
| **Precipitation** | | |
| 95th percentile | A measure of very extreme events: 95th percentile of daily precipitation | Bias (mm/day) |
| 90th percentile | A measure of extreme events: 90th percentile of daily precipitation | Bias (mm/day) |
| 50th percentile | 50th percentile of daily precipitation | Bias (mm/day) |
| 25th percentile | 25th percentile of daily precipitation | Bias (mm/day) |
| [a] Wet spell length | Mean wet spell length for a given month of the year | Bias (days) |
| [a] Dry spell length | Mean dry spell length for a given month of the year | Bias (days) |
| [a] Annual mean precipitation | Annual accumulated precipitation | Mean percentage error |
| [a] Monthly mean precipitation | Accumulated precipitation for a given month of the year | Mean percentage error |
| [b] Relative daily MSE | Mean daily square error, shown as ratio to the largest MSE result (considering both corrected and uncorrected RCMS) | MSE (ratio) |
| [b] Spearman correlation coefficient | Spearman correlation coefficients between the daily simulated and observed time series | Index |
| [a] Maximum one day precipitation (RX1day) | Maximum one-day precipitation for a given month of the year | Mean percentage error |
| [a] Simple Daily Intensity Index (SDII) | Ratio of the annual total precipitation to the number of wet days ($\geq$1 mm) in all years | Index |
| [a] Number of heavy precipitation days (R10) | Mean number of days with precipitation $\geq$ 10mm within a year | Bias (days) |
| [a] Number of very heavy precipitation days (R20) | Mean number of days with precipitation $\geq$ 20mm within a year | Bias (days) |
| [a] Very wet days (R95p) | Mean annual accumulated precipitation from days > 95th percentile in all years | Mean percentage error |
| **Temperature** | | |
| [a] Annual mean temperature | Annual mean temperature over the validation period | Mean percentage error |
| [a] Monthly mean temperature | Monthly mean temperature | Mean percentage error |
| 99th percentile of daily mean temperature | 99th percentile of the daily mean temperature | Bias (°C/day) |
| 1st percentile of daily mean temperature | 1st percentile of the daily mean temperature | Bias (°C/day) |
| [b] Pearson correlation coefficient | Pearson correlation coefficient between the daily RCM and observation time series | Index |
| **River Flow** | | |
| Q10 | A measure of high flows: river flow that is exceeded for 10% of the daily river flow time series | Bias (m$^3$/s) |
| Q95 | A measure of low flows: river flow that is exceeded for 95% of the daily river flow time series | Bias (m$^3$/s) |
| [a] Annual Q10 frequency | Mean number of days for which the observed Q10 is exceeded within a year | Bias (days) |
| [a] Annual mean river flow | Annual mean daily river flow over the validation period | Mean percentage error |
| [a] Winter (DJF) mean river flow | Winter mean daily river flow over the validation period | Mean percentage error |
| [a] Spring (MAM) mean river flow | Spring mean daily river flow over the validation period | Mean percentage error |
| [a] Summer (JJA) mean river flow | Summer mean daily river flow over the validation period | Mean percentage error |
| [a] Autumn (SON) mean river flow | Autumn mean daily river flow over the validation period | Mean percentage error |
| [b] Monthly NSE | Monthly Nash Sutcliffe Efficiency index | Index |
| [b] Relative daily MSE | Mean daily square error, shown as ratio to the largest MSE result (considering both corrected and uncorrected RCMS) | MSE (ratio) |
| [b] Spearman correlation coefficient | Spearman correlation coefficient between the daily simulated and observed time series | Index |

[a] Estimated using the long term mean (one value over the entire series)
[b] Estimated considering the time series values (one value per time step)

Table 4. Indices from the calibration and validation of the hydrological models

| Catchment | Step | Period | Daily NSE | Q10 bias | | Q95 bias | |
|---|---|---|---|---|---|---|---|
| | | | | (m³/s) | (%) | (m³/s) | (%) |
| Upper Thames | Calibration | 1986-2010 | 0.70 | -2.1 | -6 | -0.45 | -25 |
| | Validation | 1961-1985 | 0.57 | 1.5 | 5 | -0.90 | -44 |
| Glaslyn | Calibration | 1991-2010 | 0.78 | 1.0 | 8 | -0.07 | -11 |
| | Validation | 1971-1990 | 0.78 | 0.7 | 5 | -0.03 | -6 |
| Calder | Calibration | 1994-2010 | 0.62 | 1.5 | 8 | -0.31 | -16 |
| | Validation | 1976-1993 | 0.60 | 1.3 | 7 | -0.24 | -12 |
| Coquet | Calibration | 1992-2010 | 0.63 | 1.3 | 11 | -0.24 | -27 |
| | Validation | 1973-1991 | 0.52 | -0.6 | -5 | -0.25 | -31 |

Table 5. RCM rank for the temperature indices for each catchment: 1 = best, 10 = worst. The asterisks (*) indicate the resolution with the best simulation skill of each RCM in each catchment

| | | 99th percentile | 1st percentile | Annual mean | Monthly mean | Correlation | Average score | Ranking | |
|---|---|---|---|---|---|---|---|---|---|
| Upper Thames | 0.11°CCLM | 10 | 7 | 2 | 9 | 1 | 5.8 | 6 | * |
| | 0.11°HIRHAM | 3 | 9 | 3 | 5 | 6 | 5.2 | 5 | |
| | 0.11°RACMO | 2 | 8 | 9 | 7 | 4 | 6.0 | 7 | |
| | 0.11°RCA | 7 | 5 | 10 | 10 | 5 | 7.4 | 10 | |
| | 0.11°WRF | 4 | 1 | 5 | 4 | 8 | 4.4 | 2 | * |
| | 0.44°CCLM | 9 | 10 | 1 | 8 | 2 | 6.0 | 7 | |
| | 0.44°HIRHAM | 1 | 6 | 4 | 3 | 9 | 4.6 | 3 | * |
| | 0.44°RACMO | 5 | 4 | 7 | 2 | 3 | 4.2 | 1 | * |
| | 0.44°RCA | 8 | 2 | 6 | 1 | 7 | 4.8 | 4 | * |
| | 0.44°WRF | 6 | 3 | 8 | 6 | 10 | 6.6 | 9 | |
| Glaslyn | 0.11°CCLM | 9 | 2 | 4 | 3 | 1 | 3.8 | 3 | * |
| | 0.11°HIRHAM | 7 | 6 | 2 | 4 | 7 | 5.2 | 5 | * |
| | 0.11°RACMO | 3 | 7 | 1 | 1 | 4 | 3.2 | 1 | * |
| | 0.11°RCA | 2 | 4 | 3 | 2 | 6 | 3.4 | 2 | * |
| | 0.11°WRF | 4 | 8 | 5 | 6 | 10 | 6.6 | 7 | * |
| | 0.44°CCLM | 10 | 1 | 6 | 5 | 2 | 4.8 | 4 | |
| | 0.44°HIRHAM | 8 | 3 | 8 | 7 | 9 | 7.0 | 8 | |
| | 0.44°RACMO | 5 | 5 | 7 | 8 | 3 | 5.6 | 6 | |
| | 0.44°RCA | 6 | 9 | 9 | 9 | 5 | 7.6 | 9 | |
| | 0.44°WRF | 1 | 10 | 10 | 10 | 8 | 7.8 | 10 | |
| Calder | 0.11°CCLM | 9 | 7 | 8 | 8 | 1 | 6.6 | 7 | |
| | 0.11°HIRHAM | 5 | 9 | 7 | 7 | 5 | 6.6 | 7 | |
| | 0.11°RACMO | 8 | 10 | 10 | 10 | 4 | 8.4 | 9 | |
| | 0.11°RCA | 10 | 8 | 9 | 9 | 6 | 8.4 | 9 | |
| | 0.11°WRF | 7 | 3 | 1 | 4 | 8 | 4.6 | 4 | * |
| | 0.44°CCLM | 6 | 6 | 6 | 5 | 2 | 5 | 5 | * |
| | 0.44°HIRHAM | 4 | 2 | 2 | 1 | 9 | 3.6 | 2 | * |
| | 0.44°RACMO | 2 | 4 | 5 | 2 | 3 | 3.2 | 1 | * |
| | 0.44°RCA | 3 | 1 | 4 | 3 | 7 | 3.6 | 2 | * |
| | 0.44°WRF | 1 | 5 | 3 | 6 | 10 | 5 | 5 | |
| Coquet | 0.11°CCLM | 9 | 2 | 2 | 3 | 2 | 3.6 | 3 | * |
| | 0.11°HIRHAM | 1 | 3 | 3 | 2 | 5 | 2.8 | 1 | * |
| | 0.11°RACMO | 3 | 7 | 9 | 7 | 4 | 6.0 | 5 | * |
| | 0.11°RCA | 7 | 6 | 8 | 4 | 6 | 6.2 | 6 | * |
| | 0.11°WRF | 5 | 1 | 1 | 1 | 8 | 3.2 | 2 | * |
| | 0.44°CCLM | 4 | 4 | 7 | 5 | 1 | 4.2 | 4 | |
| | 0.44°HIRHAM | 10 | 8 | 5 | 6 | 9 | 7.6 | 9 | |
| | 0.44°RACMO | 6 | 9 | 6 | 9 | 3 | 6.6 | 8 | |
| | 0.44°RCA | 2 | 5 | 10 | 8 | 7 | 6.4 | 7 | |
| | 0.44°WRF | 8 | 10 | 4 | 10 | 10 | 8.4 | 10 | |

Table 6. RCM rank for the precipitation indices for each catchment: 1 = best, 10 = worst. The asterisks (*) indicate the resolution with the best simulation skill of each RCM in each catchment

| Catchment | RCM | Pr 95th | Pr 90th | Pr 50th | Pr 25th | Annual Mean | Monthly MSE | Dry Spell Length | Wet Spell Length | Monthly Mean | Correlation | SDII | R10 | R20 | R95p | RX1day | Average score | Ranking | |
|---|---|---|---|---|---|---|---|---|---|---|---|---|---|---|---|---|---|---|---|
| Upper Thames | 0.11°CCLM | 8 | 5 | 1 | 2 | 5 | 2 | 1 | 4 | 5 | 1 | 6 | 8 | 4 | 8 | 2 | 4.1 | 1 | * |
| | 0.11°HIRHAM | 7 | 4 | 4 | 3 | 3 | 5 | 6 | 6 | 1 | 3 | 5 | 7 | 6 | 3 | 7 | 4.7 | 4 | |
| | 0.11°RACMO | 3 | 2 | 9 | 8 | 7 | 3 | 4 | 5 | 4 | 10 | 9 | 3 | 5 | 5 | 9 | 5.7 | 8 | |
| | 0.11°RCA | 10 | 10 | 10 | 10 | 10 | 10 | 10 | 10 | 10 | 6 | 2 | 10 | 10 | 10 | 8 | 9.1 | 10 | |
| | 0.11°WRF | 1 | 1 | 6 | 7 | 6 | 8 | 7 | 3 | 8 | 5 | 7 | 1 | 3 | 1 | 3 | 4.5 | 2 | * |
| | 0.44°CCLM | 9 | 9 | 2 | 1 | 8 | 4 | 3 | 8 | 7 | 2 | 4 | 9 | 8 | 9 | 1 | 5.6 | 7 | |
| | 0.44°HIRHAM | 5 | 6 | 3 | 5 | 2 | 7 | 5 | 9 | 3 | 7 | 3 | 4 | 1 | 4 | 5 | 4.6 | 3 | * |
| | 0.44°RACMO | 4 | 3 | 5 | 6 | 4 | 1 | 2 | 2 | 2 | 9 | 8 | 5 | 7 | 6 | 6 | 4.7 | 5 | * |
| | 0.44°RCA | 2 | 8 | 8 | 4 | 9 | 9 | 9 | 1 | 9 | 4 | 1 | 2 | 1 | 2 | 4 | 4.9 | 6 | * |
| | 0.44°WRF | 6 | 7 | 7 | 9 | 1 | 6 | 8 | 7 | 6 | 8 | 10 | 6 | 9 | 7 | 10 | 7.1 | 9 | |
| Glaslyn | 0.11°CCLM | 5 | 5 | 8 | 2 | 5 | 5 | 6 | 5 | 5 | 1 | 5 | 5 | 5 | 5 | 5 | 4.8 | 5 | * |
| | 0.11°HIRHAM | 1 | 1 | 6 | 5 | 1 | 3 | 5 | 3 | 2 | 3 | 1 | 3 | 1 | 1 | 1 | 2.5 | 1 | * |
| | 0.11°RACMO | 3 | 3 | 3 | 9 | 3 | 1 | 3 | 8 | 3 | 2 | 3 | 2 | 3 | 2 | 4 | 3.5 | 3 | * |
| | 0.11°RCA | 2 | 2 | 2 | 10 | 2 | 2 | 8 | 6 | 1 | 6 | 2 | 1 | 2 | 3 | 2 | 3.4 | 2 | * |
| | 0.11°WRF | 4 | 4 | 1 | 6 | 4 | 4 | 4 | 4 | 4 | 7 | 4 | 4 | 4 | 4 | 3 | 4.1 | 4 | * |
| | 0.44°CCLM | 10 | 9 | 10 | 3 | 9 | 9 | 9 | 9 | 9 | 5 | 9 | 9 | 9 | 8 | 7 | 8.3 | 9 | |
| | 0.44°HIRHAM | 9 | 10 | 9 | 1 | 10 | 10 | 10 | 10 | 10 | 9 | 7 | 10 | 10 | 9 | 9 | 8.9 | 10 | |
| | 0.44°RACMO | 7 | 7 | 4 | 7 | 7 | 7 | 2 | 1 | 7 | 4 | 10 | 7 | 7 | 7 | 8 | 6.1 | 7 | |
| | 0.44°RCA | 8 | 8 | 7 | 4 | 8 | 8 | 7 | 7 | 8 | 8 | 8 | 8 | 10 | 10 | 10 | 7.8 | 8 | |
| | 0.44°WRF | 6 | 6 | 5 | 8 | 6 | 6 | 1 | 2 | 6 | 10 | 6 | 6 | 6 | 6 | 6 | 5.7 | 6 | |
| Calder | 0.11°CCLM | 1 | 2 | 2 | 1 | 1 | 1 | 2 | 3 | 1 | 1 | 1 | 2 | 2 | 2 | 9 | 2.1 | 1 | * |
| | 0.11°HIRHAM | 10 | 10 | 8 | 5 | 9 | 9 | 7 | 8 | 9 | 5 | 7 | 9 | 10 | 10 | 10 | 8.4 | 9 | |
| | 0.11°RACMO | 2 | 1 | 9 | 9 | 3 | 5 | 5 | 9 | 4 | 4 | 4 | 1 | 1 | 1 | 3 | 4.1 | 2 | * |
| | 0.11°RCA | 9 | 9 | 10 | 10 | 10 | 10 | 10 | 10 | 10 | 6 | 3 | 10 | 9 | 9 | 1 | 8.4 | 10 | |
| | 0.11°WRF | 3 | 3 | 6 | 4 | 6 | 8 | 6 | 5 | 7 | 8 | 2 | 4 | 3 | 3 | 5 | 5.1 | 5 | * |
| | 0.44°CCLM | 6 | 7 | 4 | 2 | 8 | 3 | 4 | 4 | 8 | 2 | 5 | 7 | 4 | 6 | 2 | 4.8 | 3 | |
| | 0.44°HIRHAM | 4 | 4 | 1 | 3 | 7 | 4 | 9 | 6 | 6 | 7 | 6 | 3 | 5 | 4 | 5 | 4.9 | 4 | * |
| | 0.44°RACMO | 8 | 8 | 7 | 7 | 5 | 2 | 3 | 1 | 3 | 3 | 10 | 8 | 8 | 8 | 6 | 5.8 | 7 | |
| | 0.44°RCA | 7 | 6 | 3 | 6 | 4 | 7 | 8 | 2 | 5 | 9 | 8 | 6 | 7 | 7 | 7 | 6.1 | 8 | * |
| | 0.44°WRF | 5 | 5 | 5 | 8 | 2 | 6 | 1 | 7 | 2 | 10 | 9 | 5 | 6 | 5 | 4 | 5.3 | 6 | |
| Coquet | 0.11°CCLM | 4 | 5 | 1 | 1 | 2 | 1 | 1 | 3 | 1 | 1 | 3 | 4 | 2 | 1 | 2 | 2.1 | 1 | * |
| | 0.11°HIRHAM | 6 | 9 | 9 | 7 | 9 | 9 | 9 | 7 | 9 | 5 | 1 | 7 | 5 | 6 | 4 | 6.8 | 8 | |
| | 0.11°RACMO | 5 | 3 | 6 | 8 | 1 | 3 | 7 | 5 | 2 | 4 | 9 | 5 | 4 | 5 | 5 | 4.8 | 4 | * |
| | 0.11°RCA | 10 | 10 | 10 | 10 | 10 | 10 | 10 | 10 | 10 | 9 | 7 | 10 | 9 | 10 | 1 | 9.1 | 10 | |
| | 0.11°WRF | 2 | 1 | 5 | 3 | 3 | 6 | 5 | 6 | 3 | 7 | 5 | 2 | 1 | 2 | 3 | 3.6 | 2 | * |
| | 0.44°CCLM | 7 | 6 | 4 | 2 | 8 | 4 | 3 | 8 | 7 | 2 | 3 | 6 | 7 | 7 | 6 | 5.3 | 6 | |
| | 0.44°HIRHAM | 3 | 2 | 8 | 9 | 4 | 5 | 8 | 2 | 4 | 8 | 1 | 1 | 3 | 4 | 8 | 4.7 | 3 | * |
| | 0.44°RACMO | 8 | 7 | 3 | 4 | 6 | 2 | 2 | 4 | 6 | 3 | 9 | 8 | 10 | 9 | 10 | 6.1 | 7 | |
| | 0.44°RCA | 1 | 4 | 7 | 5 | 5 | 8 | 6 | 1 | 5 | 6 | 7 | 3 | 6 | 3 | 7 | 4.9 | 5 | * |
| | 0.44°WRF | 9 | 8 | 2 | 6 | 7 | 7 | 4 | 9 | 8 | 10 | 5 | 9 | 8 | 8 | 9 | 7.3 | 9 | |

Table 7. RCM rank for the river flow indices for each catchment: 1 = best, 10 = worst. The asterisks (*) indicate the resolution with the best simulation skill of each RCM in each catchment

| Catchment | | Annual mean | Winter mean | Spring mean | Summer mean | Autumn mean | Monthly NSE | Daily MSE | Spearman correl. | Q10 | Annual Q10 events | Q95 | Average | Rank | |
|---|---|---|---|---|---|---|---|---|---|---|---|---|---|---|---|
| Upper Thames | 0.11° CCLM | 1 | 2 | 2 | 3 | 3 | 1 | 1 | 4 | 1 | 1 | 1 | 1.8 | 1 | * |
| | 0.11° HIRHAM | 3 | 4 | 4 | 2 | 1 | 3 | 3 | 1 | 3 | 3 | 4 | 2.8 | 2 | * |
| | 0.11° RACMO | 8 | 9 | 8 | 6 | 9 | 9 | 8 | 7 | 8 | 8 | 9 | 8.1 | 9 | |
| | 0.11° RCA | 10 | 10 | 10 | 10 | 10 | 10 | 10 | 10 | 10 | 10 | 10 | 10.0 | 10 | |
| | 0.11° WRF | 7 | 1 | 6 | 9 | 5 | 6 | 6 | 8 | 5 | 6 | 8 | 6.1 | 6 | * |
| | 0.44° CCLM | 4 | 7 | 1 | 4 | 7 | 2 | 2 | 3 | 6 | 4 | 3 | 3.9 | 4 | |
| | 0.44° HIRHAM | 2 | 5 | 3 | 1 | 2 | 4 | 5 | 5 | 2 | 2 | 2 | 3.0 | 3 | |
| | 0.44° RACMO | 5 | 6 | 5 | 5 | 6 | 5 | 4 | 2 | 4 | 5 | 7 | 4.9 | 5 | * |
| | 0.44° RCA | 9 | 8 | 9 | 7 | 8 | 8 | 9 | 6 | 9 | 9 | 6 | 8.0 | 8 | * |
| | 0.44° WRF | 6 | 3 | 7 | 8 | 4 | 7 | 7 | 9 | 7 | 7 | 5 | 6.4 | 7 | |
| Glaslyn | 0.11° CCLM | 5 | 5 | 5 | 6 | 6 | 5 | 5 | 4 | 5 | 5 | 8 | 5.36 | 5 | * |
| | 0.11° HIRHAM | 1 | 1 | 1 | 4 | 1 | 1 | 3 | 2 | 1 | 1 | 4 | 1.82 | 1 | * |
| | 0.11° RACMO | 2 | 2 | 2 | 3 | 3 | 2 | 1 | 1 | 2 | 2 | 2 | 2 | 2 | * |
| | 0.11° RCA | 3 | 3 | 3 | 1 | 2 | 3 | 2 | 6 | 3 | 3 | 1 | 2.73 | 3 | * |
| | 0.11° WRF | 4 | 4 | 4 | 2 | 4 | 4 | 4 | 5 | 4 | 4 | 3 | 3.82 | 4 | * |
| | 0.44° CCLM | 9 | 9 | 10 | 10 | 10 | 9 | 9 | 8 | 10 | 8 | 9 | 9.18 | 9 | |
| | 0.44° HIRHAM | 10 | 10 | 9 | 9 | 9 | 10 | 10 | 9 | 9 | 8 | 10 | 9.36 | 10 | |
| | 0.44° RACMO | 7 | 6 | 7 | 8 | 7 | 7 | 7 | 3 | 7 | 7 | 6 | 6.55 | 7 | |
| | 0.44° RCA | 8 | 8 | 8 | 7 | 8 | 8 | 8 | 10 | 8 | 10 | 7 | 8.18 | 8 | |
| | 0.44° WRF | 6 | 7 | 6 | 5 | 5 | 6 | 6 | 7 | 6 | 6 | 5 | 5.91 | 6 | |
| Calder | 0.11° CCLM | 2 | 2 | 6 | 4 | 5 | 2 | 1 | 2 | 1 | 1 | 1 | 2.45 | 2 | * |
| | 0.11° HIRHAM | 9 | 10 | 9 | 8 | 9 | 9 | 9 | 4 | 9 | 9 | 9 | 8.55 | 9 | |
| | 0.11° RACMO | 6 | 6 | 7 | 6 | 6 | 5 | 6 | 3 | 6 | 8 | 8 | 6.09 | 6 | |
| | 0.11° RCA | 10 | 9 | 10 | 10 | 10 | 10 | 10 | 9 | 10 | 10 | 10 | 9.82 | 10 | |
| | 0.11° WRF | 7 | 8 | 8 | 9 | 2 | 8 | 8 | 6 | 7 | 7 | 7 | 7 | 8 | |
| | 0.44° CCLM | 8 | 5 | 5 | 7 | 8 | 6 | 3 | 5 | 8 | 6 | 6 | 6.09 | 6 | |
| | 0.44° HIRHAM | 5 | 4 | 3 | 5 | 7 | 4 | 5 | 10 | 5 | 5 | 4 | 5.18 | 5 | * |
| | 0.44° RACMO | 3 | 3 | 2 | 1 | 1 | 1 | 2 | 1 | 3 | 2 | 2 | 1.91 | 1 | * |
| | 0.44° RCA | 4 | 7 | 4 | 3 | 3 | 7 | 7 | 8 | 4 | 4 | 5 | 5.09 | 4 | * |
| | 0.44° WRF | 1 | 1 | 1 | 2 | 4 | 3 | 4 | 7 | 2 | 3 | 3 | 2.82 | 3 | * |
| Coquet | 0.11° CCLM | 1 | 5 | 1 | 1 | 6 | 4 | 5 | 1 | 4 | 1 | 1 | 2.73 | 1 | * |
| | 0.11° HIRHAM | 9 | 6 | 9 | 9 | 8 | 9 | 9 | 5 | 8 | 9 | 9 | 8.18 | 9 | |
| | 0.11° RACMO | 7 | 1 | 7 | 7 | 5 | 6 | 3 | 6 | 2 | 2 | 8 | 4.91 | 5 | |
| | 0.11° RCA | 10 | 10 | 10 | 10 | 10 | 10 | 10 | 10 | 10 | 10 | 10 | 10 | 10 | |
| | 0.11° WRF | 2 | 2 | 2 | 5 | 2 | 2 | 6 | 4 | 1 | 3 | 5 | 3.09 | 2 | * |
| | 0.44° CCLM | 8 | 9 | 8 | 4 | 9 | 7 | 2 | 2 | 9 | 8 | 2 | 6.18 | 7 | |
| | 0.44° HIRHAM | 6 | 3 | 4 | 8 | 7 | 8 | 8 | 9 | 5 | 7 | 7 | 6.55 | 8 | * |
| | 0.44° RACMO | 3 | 7 | 3 | 2 | 3 | 1 | 1 | 3 | 6 | 6 | 4 | 3.55 | 3 | * |
| | 0.44° RCA | 5 | 4 | 6 | 6 | 4 | 5 | 7 | 7 | 3 | 4 | 6 | 5.18 | 6 | * |
| | 0.44° WRF | 4 | 8 | 5 | 3 | 1 | 3 | 4 | 8 | 7 | 5 | 3 | 4.64 | 4 | |

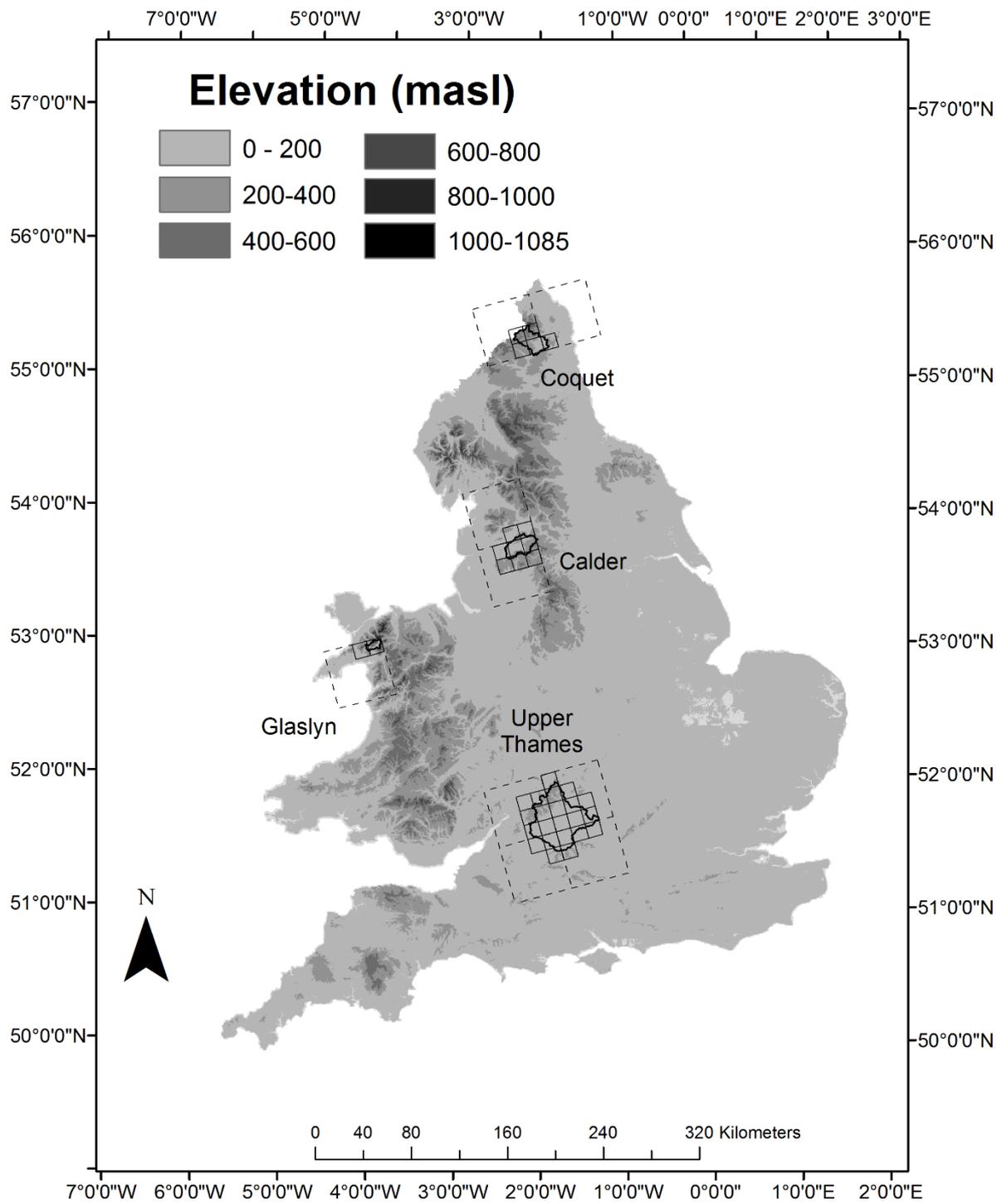

Figure 1. Location of the study catchments and the RCM grid boxes used for their simulation. The 0.11° and 0.44° grid boxes are shown with solid and dashed lines, respectively

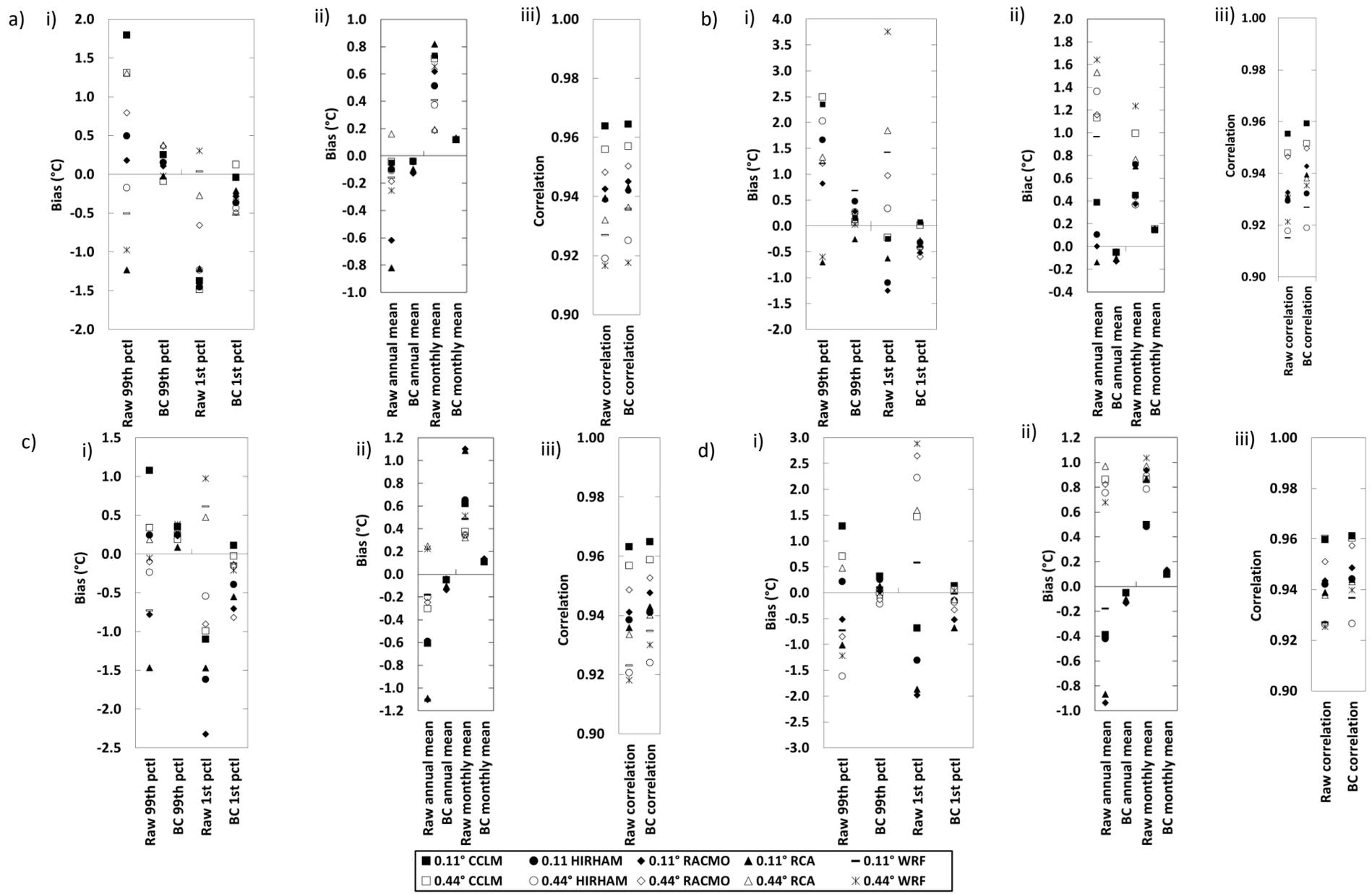

Figure 2. Results of the temperature performance measures, described on Table 3, for the a) upper Thames, b) Glaslyn, c) Calder and d) Coquet catchments. Please note the differences in the y-axis (BC = Bias corrected)

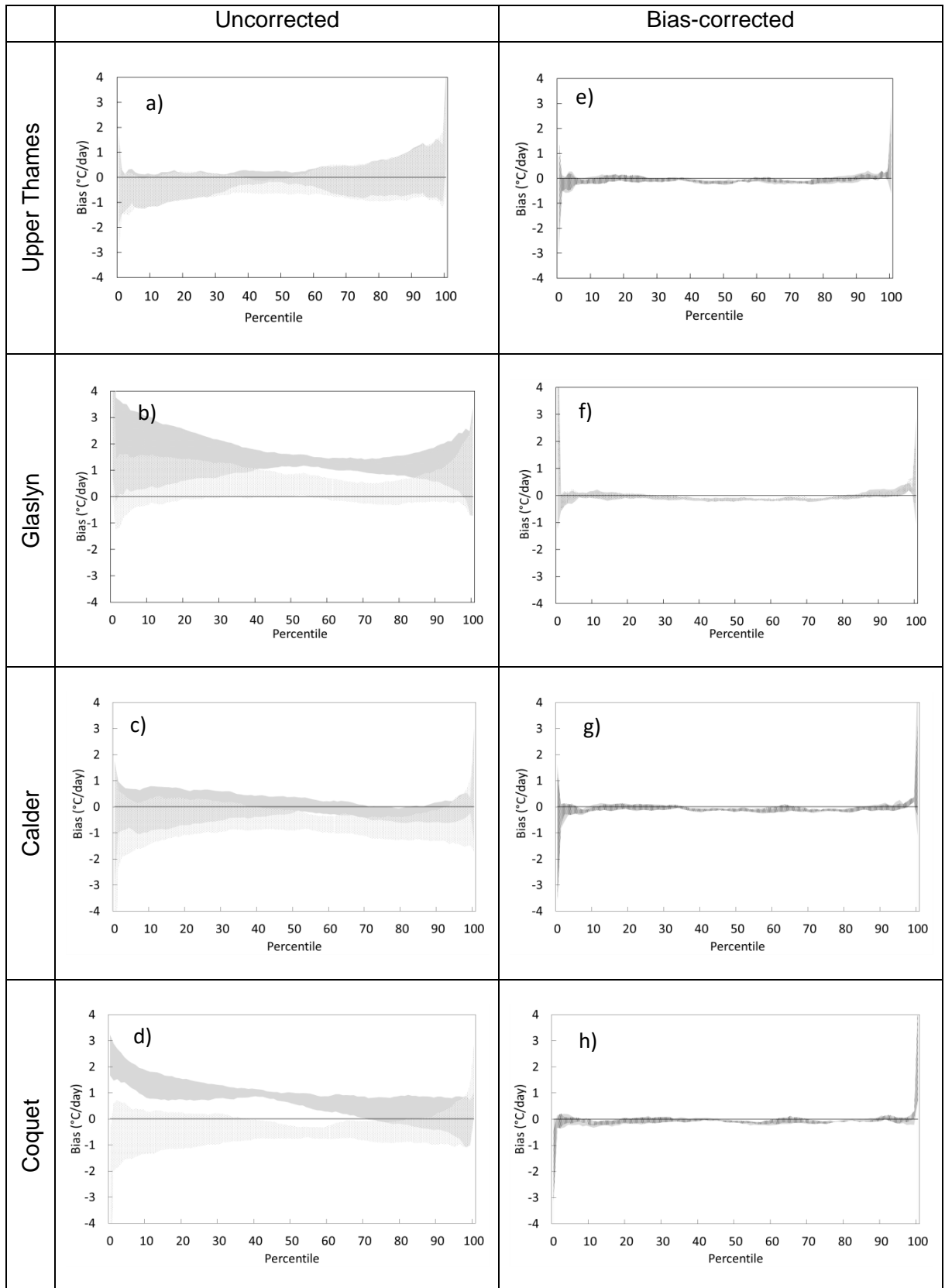

Figure 3. Temperature percentile biases for the uncorrected and bias-corrected RCMs. The solid fill represents the spread form the 0.44° RCMs and the dotted fill is the spread from the 0.11° RCMs

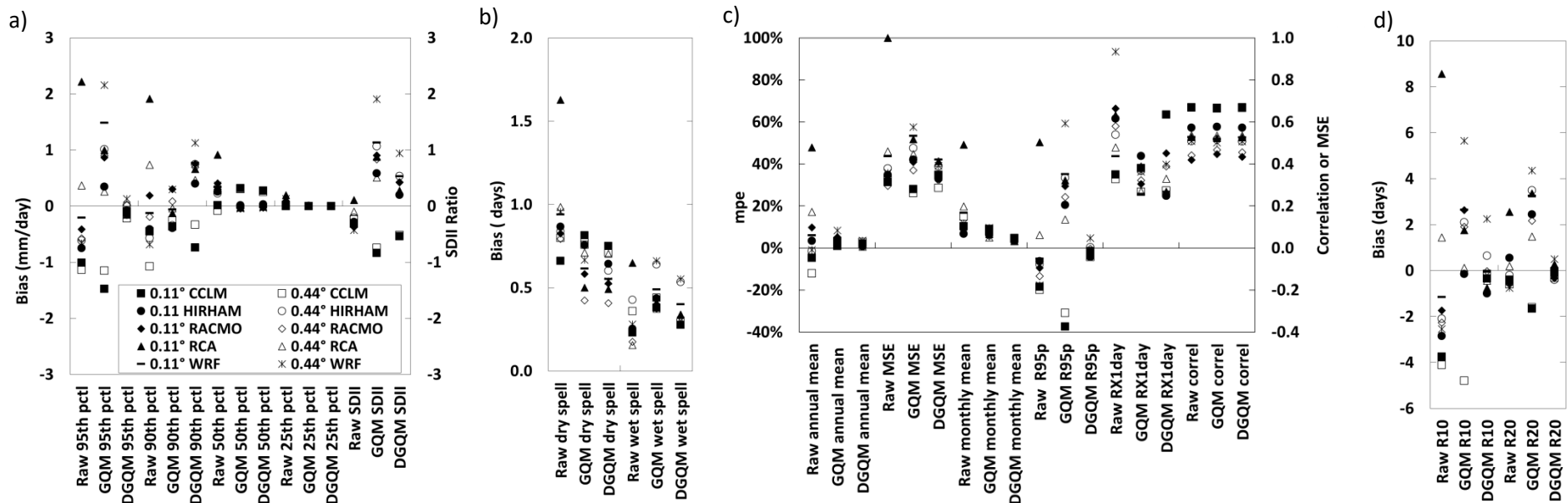

Figure 4. Results of the precipitation performance measures for the upper Thames catchment. Please note the differences in the y-axis. For definitions of the performance measures refer to Table 3 (BC-1G = Bias corrected using the Gamma distribution QM approach, BC-2G = Bias corrected using the Double Gamma distribution approach)

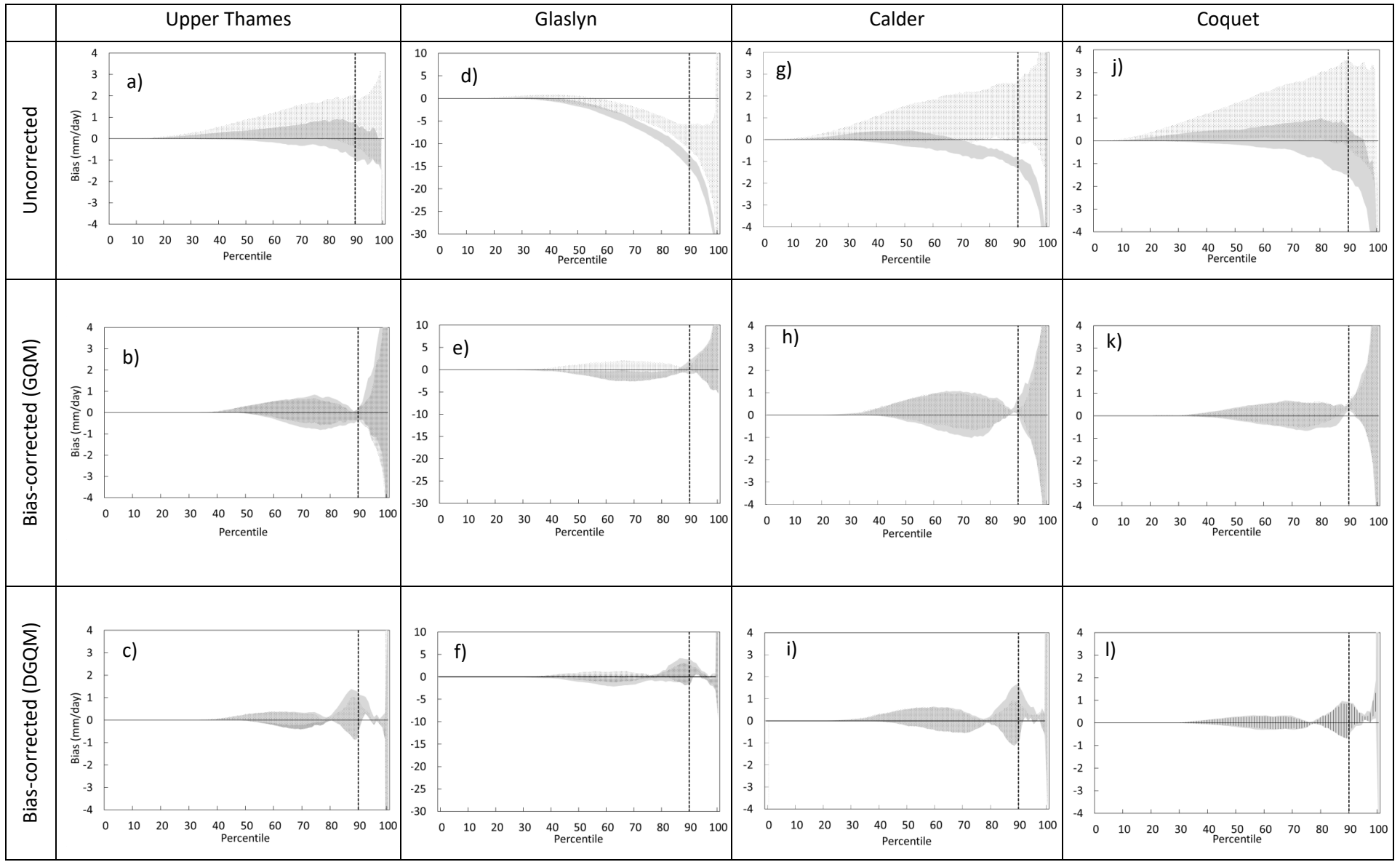

Figure 5. Precipitation percentile biases for the uncorrected and bias-corrected RCMs using the Gamma distribution (GQM) and Double Gamma distribution (DGQM) QM. The solid fill represents the spread of the 0.44° RCMs and the dotted fill the spread of the 0.11° RCMs. The 90th precipitation percentile is indicated by a vertical dotted line.

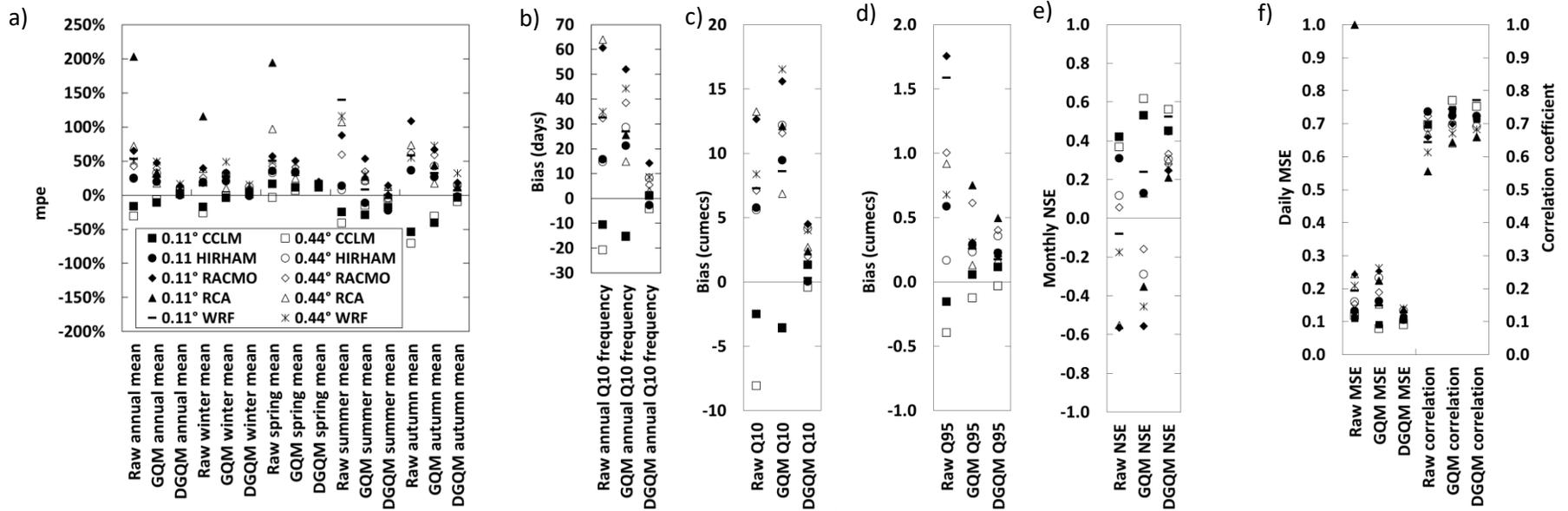

Figure 6. Results of the river flow performance measures for the upper Thames catchment. Please note the differences in the y-axis. For definitions of the performance measures refer to Table 3 (GQM = Gamma distribution Quantile Mapping and DGQM = double Gamma distribution Quantile Mapping)

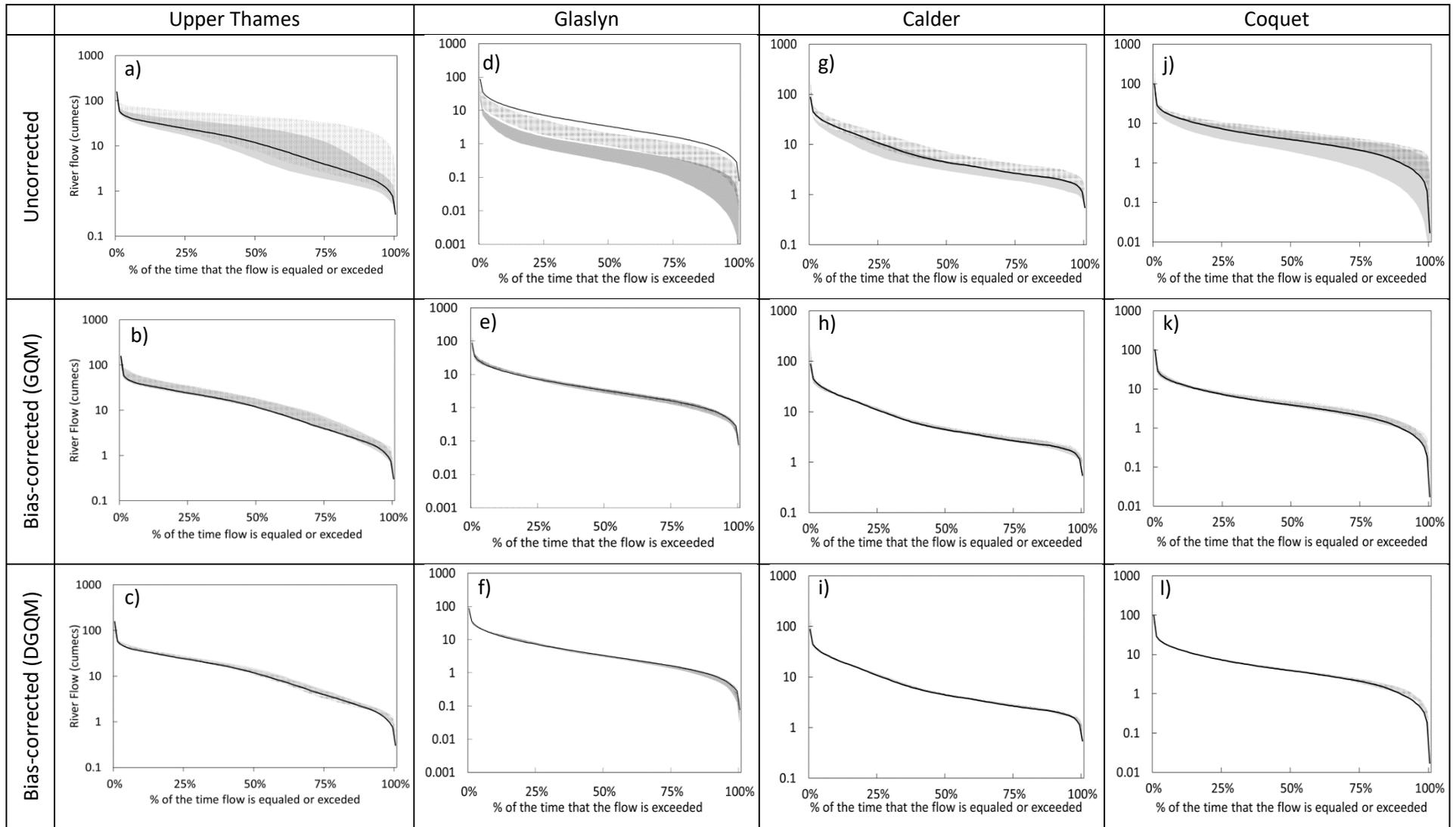

Figure 7. Flow duration curve biases from using the uncorrected and bias-corrected temperature and precipitation simulations. The 0.44° RCMs spread is shown with a solid fill, the 0.11° RCMs spread with a dotted fill and the reference FDC with a solid line. (GQM = Gamma distribution Quantile Mapping and DGQM = double Gamma distribution Quantile Mapping)

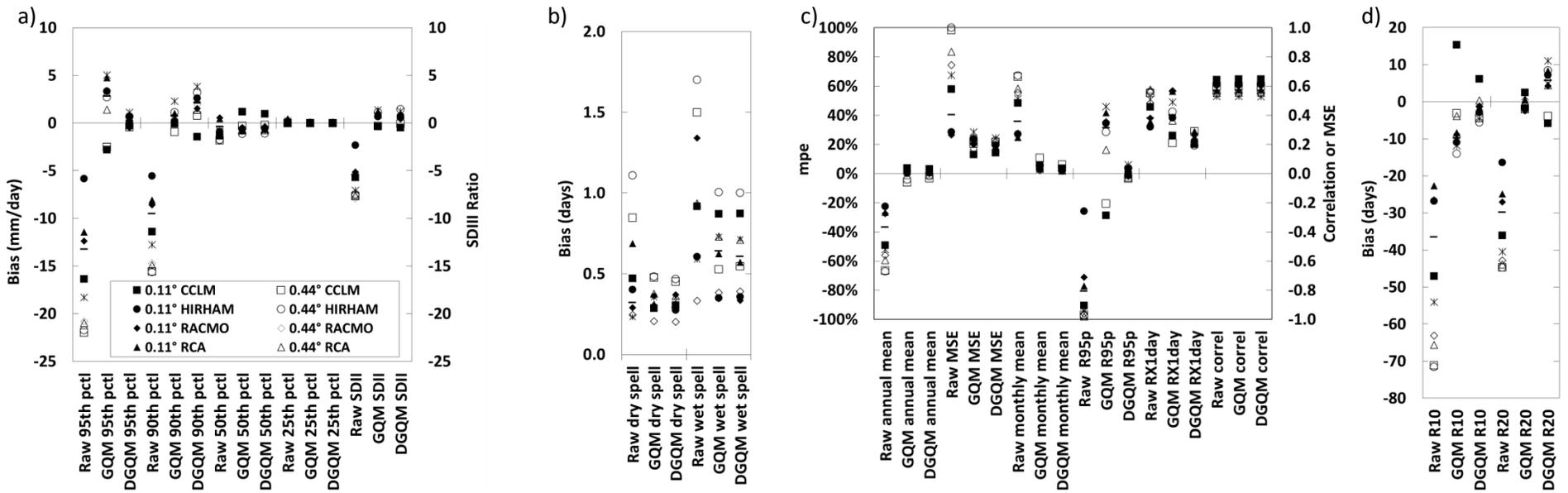

Figure S1. Similar to Figure 5 but for the Glaslyn catchment

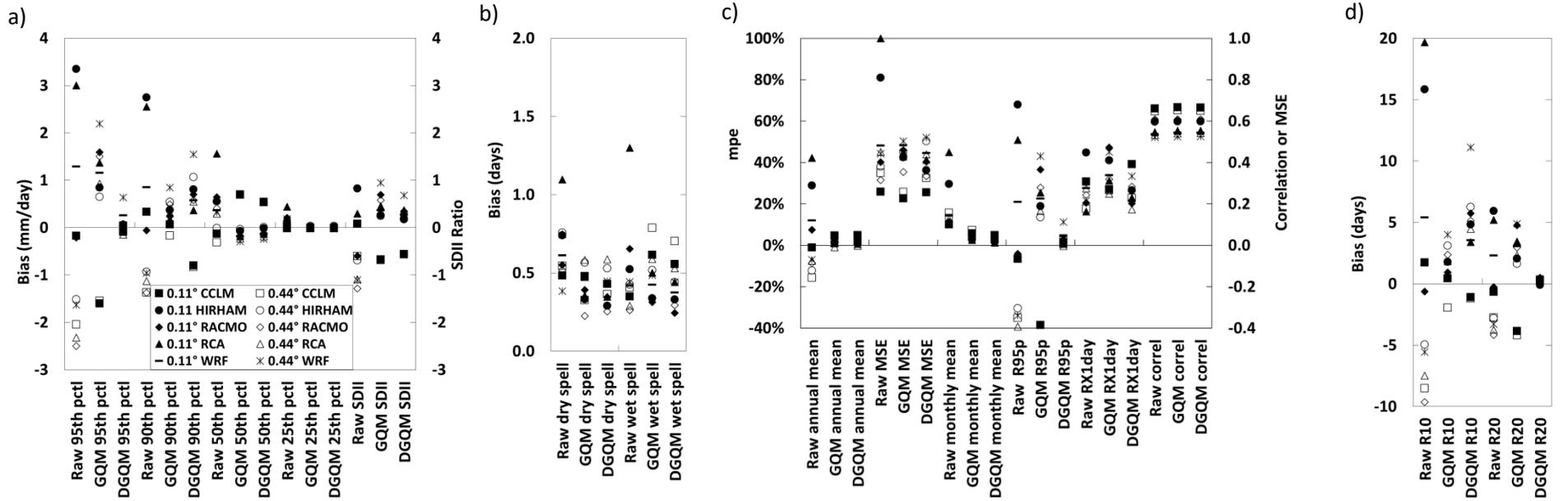

Figure S2. Similar to Figure 5 but for the Calder catchment

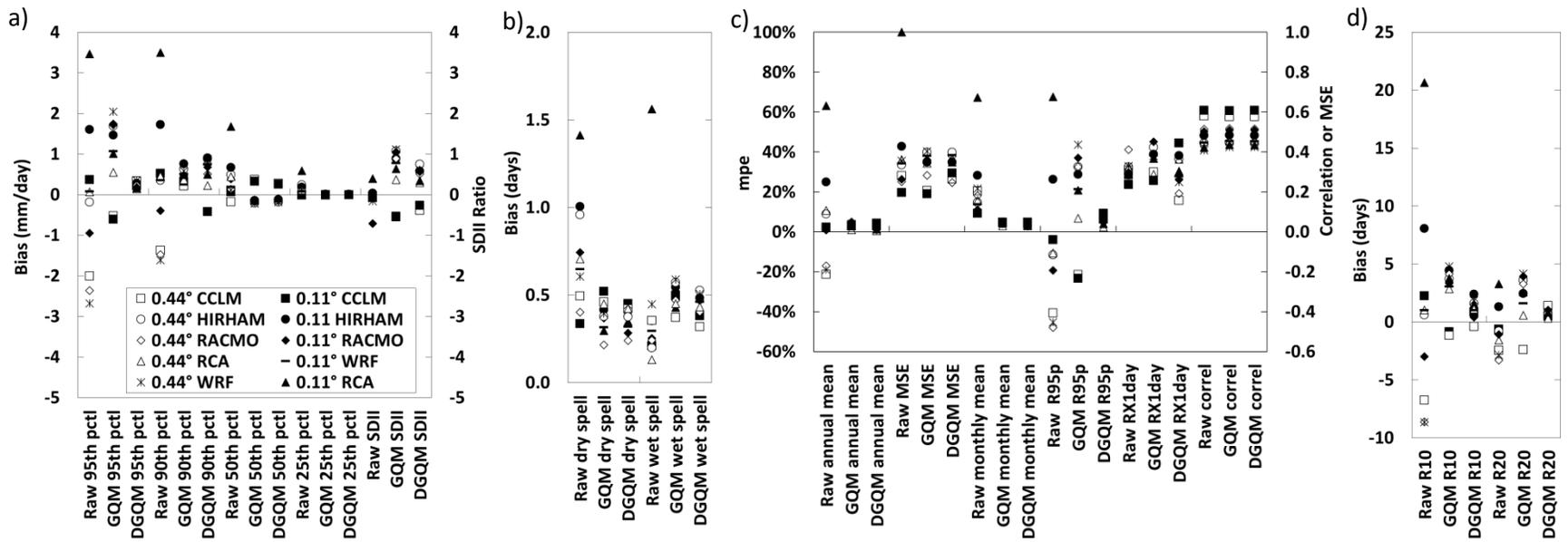

Figure S3. Similar to Figure 5 but for the Coquet catchment

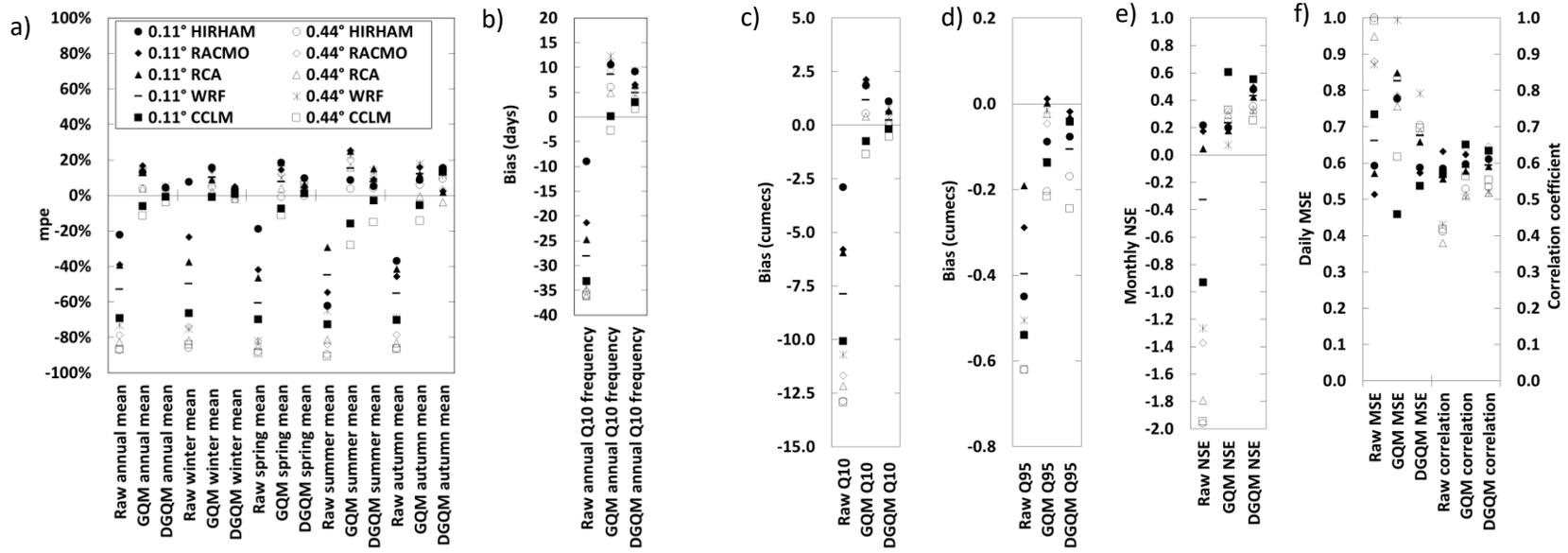

Figure S4. Similar to Figure 7 but for the Glaslyn catchment

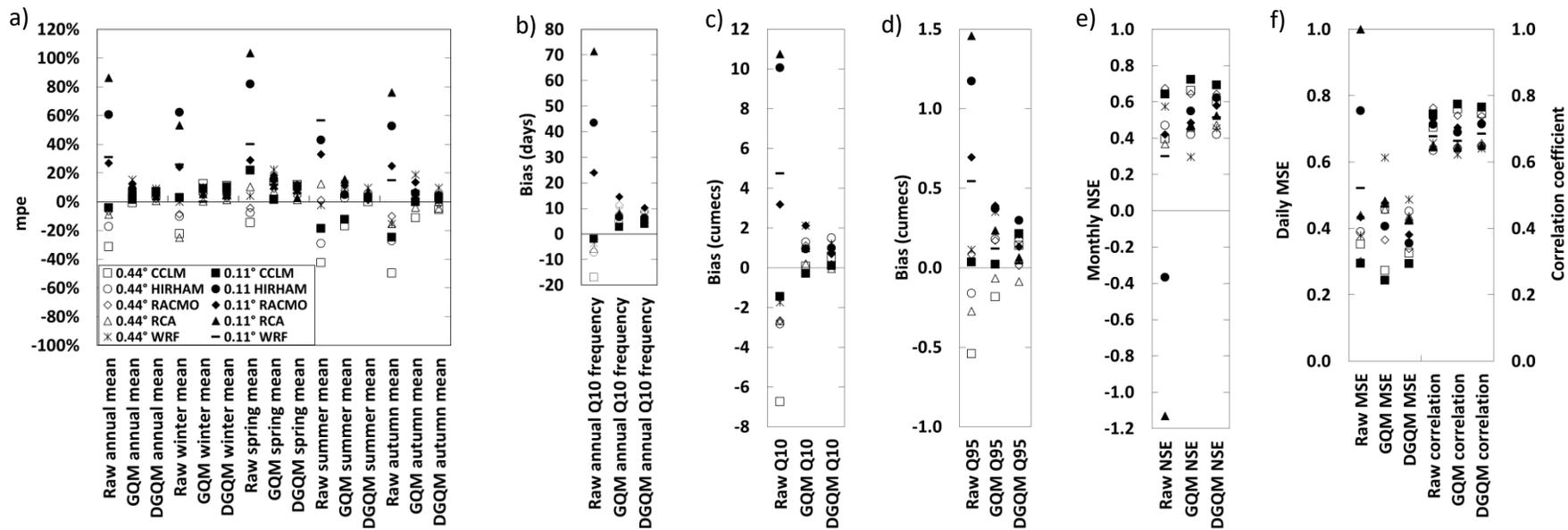

Figure S5. Similar to Figure 7 but for the Calder catchment

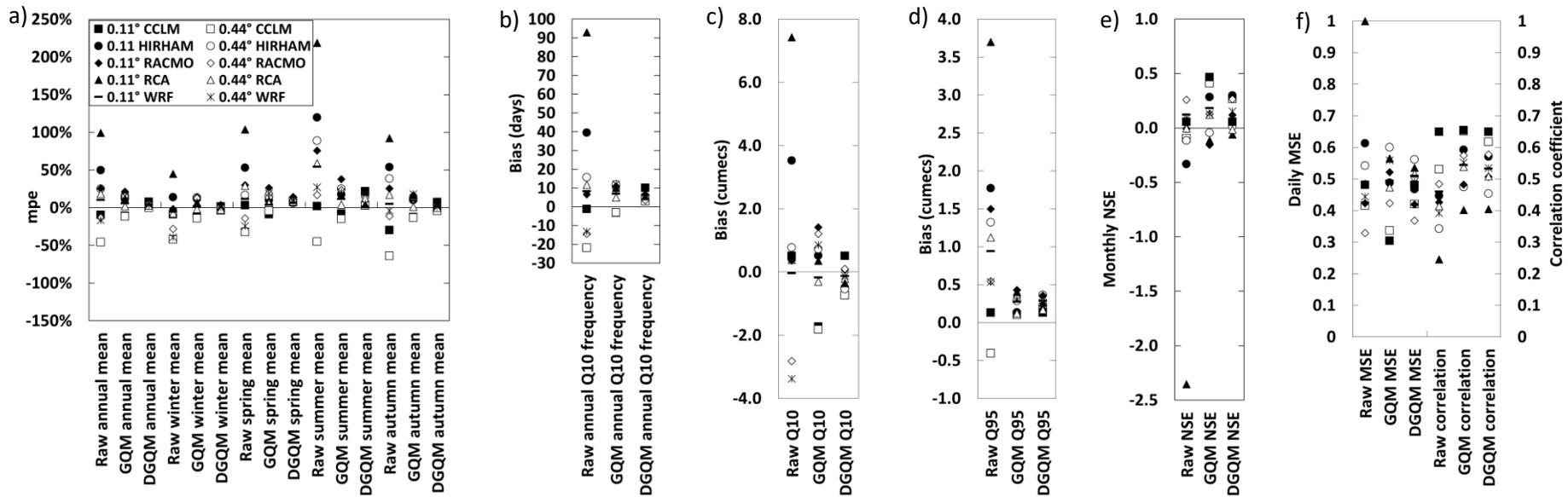

Figure S6. Similar to Figure 7 but for the Coquet catchment